\documentclass[11pt]{emulateapj}
\usepackage{natbib}
\usepackage{graphicx}
\usepackage{amsmath}
\usepackage{amssymb}

\def\jgr{J. Geophys. Res. }

\def\apj{Astrophys. J. }
\def\chf{CH$_4$ }
\def\chfx{CH$_4$}
\def\chisq{$\chi^2$ }
\def\chisqx{$\chi^2$}
\def\icm{cm$^{-1}$ }
\def\icmx{cm$^{-1}$}

\def\Wm2{W/m$^2$}
\def\Wpm2sr{Wm$^{-2}sr^{-1}$}
\def\deg{$^\circ$ }
\def\degx{$^\circ$}

\def\mum{$\mu$m }
\def\mumx{$\mu$m}

\begin{document}
\title{Methane on Uranus: The case for a compact
\chf cloud layer at low latitudes and a severe \chf depletion at
high-latitudes based on re-analysis of Voyager occultation
measurements and STIS spectroscopy\footnotemark[\dag]}
\author{L.~A. Sromovsky\altaffilmark{1},
  P.~M. Fry\altaffilmark{1},
  J.~H. Kim\altaffilmark{1}}
\altaffiltext{1}{University of Wisconsin - Madison, Madison WI 53706}
\altaffiltext{\dag}{Based in part on observations with the NASA/ESA Hubble Space
Telescope obtained at the Space Telescope Science Institute, which is
operated by the Association of Universities for Research in Astronomy,
Incorporated under NASA Contract NAS5-26555.} 

\slugcomment{Journal reference: Icarus 215 (2011) 292-312.}
\begin{abstract} 
Lindal et al. (1987, J. Geophys. Res. 92, 14987-15001) presented a
range of temperature and methane profiles for Uranus that were
consistent with 1986 Voyager radio occultation measurements of
refractivity versus altitude.  A localized refractivity slope
variation near 1.2 bars was interpreted to be the result of a
condensed methane cloud layer.  However, models fit to near-IR spectra
found particle concentrations much deeper in the atmosphere, in the
1.5-3 bar range (Sromovsky et al. 2006, Icarus 182, 577-593, Sromovsky
and Fry 2008, Icarus 193, 211-229, Irwin et al. 2010, Icarus 208,
913-926), and a recent analysis of STIS spectra argued for a model in
which aerosol particles formed diffusely distributed hazes, with no
compact condensation layer (Karkoschka and Tomasko 2009, Icarus 202,
287-309).  To try to reconcile these results, we reanalyzed the
occultation observations with the He volume mixing ratio reduced from
0.15 to 0.116, which is near the edge of the 0.033 uncertainty range
given by Conrath et al. (1987, J. Geophys. Res., 15003-10).  This
allowed us to obtain saturated mixing ratios within the putative cloud
layer and to reach above-cloud and deep methane mixing ratios
compatible with STIS spectral constraints.  Using a 5-layer vertical
aerosol model with two compact cloud layers in the 1-3 bar region, we
find that the best fit pressure for the upper layer is virtually
identical to the pressure range inferred from the occultation analysis
for a methane mixing ratio near 4\% at 5\deg S.  This strongly argues
that Uranus does indeed have a compact methane cloud layer.  In
addition, our cloud model can fit the latitudinal variations in
spectra between 30\deg S and 20\deg N, using the same profiles of
temperature and methane mixing ratio. But closer to the pole, the
model fails to provide accurate fits without introducing an
increasingly strong upper tropospheric depletion of methane at
increased latitudes, in rough agreement with the trend identified by
Karkoschka and Tomasko (2009, Icarus 202, 287-309).
\end{abstract}


\maketitle
\shortauthors{Sromovsky et al.} 
\shorttitle{Methane on Uranus: Compact cloud layer and high-latitude depletion}

\section{Introduction}

The existence of a thin methane ice cloud in the atmosphere of Uranus
was inferred by \cite{Lindal1987}, henceforth referred to as {\bf
L87}, from their analysis of Voyager 2 radio occultation measurements.
They also derived a suite of temperature and methane profiles that
were all consistent with their measurements, including Model F, which
had the greatest deep methane mixing ratio (4\%), and their preferred
Model D, which had a deep mixing ratio of 2.3\% and an above-cloud
relative humidity of 30\%. (Here we use the term relative humidity of
methane to refer to the ratio of its partial pressure to its
saturation pressure at the same temperature, and the mixing ratio
referred to is the volume mixing ratio or VMR.)  A methane cloud layer
near the 1.2-bar level inferred by L87 has been used successfully in
the analysis of observations in the visible spectral range. For
example, \cite{Rages1991} incorporated a methane cloud layer into
their models of Voyager 2 imaging observations, finding modest optical
depths of 0.66$\pm$0.18 at 22\deg S (assumed to be independent of
wavelength), and about three times that level at 65 \deg S, while
assuming a deep methane mixing ratio of 4\% (consistent with L87 Model
F).  On the other hand, an analysis of hydrogen S(0) and S(1) quadrupole
line features by \cite{Baines1995}, which also incorporated a methane
cloud near 1.2 bars, found its opacity (weighted to high latitudes) to
be about 0.4 at 0.6 \mumx, while inferring a deep methane mixing ratio of
1.6\%. Neither of these authors tried to constrain the pressure of the
cloud, however, so that consistency with occultation results was not
fully established.

Recently, a serious challenge to the existence of the methane cloud
layer was made by \cite{Kark2009IcarusSTIS}, henceforth referred to as
{\bf KT2009}, based on their analysis of spatially resolved 0.3-1 \mum
spectra obtained from 2002 observations by the Hubble Space Telescope
Imaging Spectrograph (STIS).  They concluded that the most significant
cloud opacity concentration was in a layer from 1.2-2 bars, with
particles uniformly mixed with the gas in this layer, which had
wavelength-independent optical depths between 1.2 and 2.2. They argued
for no localized \chf condensation layer at all, but instead for the
existence of a global thick and diffuse tropospheric haze similar to
that observed on Titan.  This seemed to confirm the analysis of
near-IR spectral observations, which had already questioned the
existence of a methane ice cloud near 1.2 bars.

From an analysis of the \cite{Fink1979} near-IR spectrum, which made
use of improved methane absorption coefficients
\citep{Irwin2006ch42e}, \cite{Sro2006ch4} obtained a cloud layer near
2 bars for the L87 Model D profiles and at 1.5-1.7 bars for their
Model F profiles.  A subsequent analysis of 2004 near-IR imaging
observations by \cite{Sro2007struc} concluded that the methane
relative humidity should be near 60\%-100\% above the nominal cloud
region (Models D and F had 30\% and 53\% respectively), and at low
latitudes found no need for a cloud layer in the 1.2-2 bar region.
Further refinements of methane absorption models for the near-IR
\citep{Kark2010ch4} did not entirely fix these discrepancies.
\cite{Irwin2010Icar} used these improved methane coefficients and the
L87 Model D T(P) profile and above-cloud methane mixing ratio, but
used 1.6\% instead of 2.26\% below the putative cloud layer. With
these assumptions they obtained a low-latitude cloud density peak near
2.5 bars. Even using the L87 Model F T(P) profile and a 4\% deep
methane VMR, \cite{Irwin2010Icar} still obtained a cloud peak that was
too deep to reach the methane condensation level, though the cloud
pressure was then elevated to about the 1.7-bar level, which puts the
peak in the middle of the main aerosol layer of KT2009.

More methane seems to be required to bring the cloud pressure inferred
from spectral observations to the same level as inferred from the
refractivity profile.  However, according to L87, their Model F is an
upper limit on methane amounts, and they argued that Model D is really
preferable.  That solution has a deep methane mixing ratio of 2.3\% by
volume, a cloud layer between 1.15 and 1.27 bars, and an above-cloud
methane humidity of 30\%. Other profiles that satisfy the occultation
measurements have deep methane mixing ratios varying from zero to 4\%,
and above-cloud humidities varying from zero to 53\%, while the
in-cloud humidities vary from zero to 78\%.  Model D was preferred for
three reasons: (1) it provides the best agreement with IRIS
observations sampling the above-cloud region, (2) it has the highest
in-cloud humidity levels, (3) it yields a deep mixing ratio (2.3\%)
that is in close agreement with that of \cite{Orton1986Icar}.  The
first reason is weak because the IRIS observations in question
\citep{Conrath1987JGR} are at a large zenith angle and rather
uncertain. The second is weakened by the fact that only the nominal
helium volume mixing ratio was considered, and that can strongly
affect the methane humidity, as we will show here. The
third reason is questionable because the Orton et al. temperature
profile derived assuming 2\% methane does not agree with the
occultation profile using virtually the same mixing ratio. Much
stronger external constraints are available from spectral observations
in the CCD ($\sim$0.3 - 1 \mumx) wavelength region, as shown by KT2009 and by the analysis
presented here.  In prior use of these spectral constraints however, there has
been an unjustified deviation from occultation constraints in both
thermal and methane profiles, in which a thermal profile derived for
one methane profile is used for a very different mixing ratio
\citep{Irwin2010Icar, Baines1995, Sro2007struc} and above-cloud methane
profiles have been used that exceed all of the occultation solutions,
e.g. KT2009 and \cite{Baines1995}.  KT2009 also inferred that the methane
mixing ratio varies with latitude, which raises additional
questions about the effects of corresponding density variations
with latitude.   

To summarize, where spectral observations have been used to test the
location of cloud layers on Uranus, the inferred locations are
considerably deeper than implied by the occultation observations.  And
the spectral constraints on the methane mixing ratio range from a low
of 1.6$^{+0.7}_{-0.5}$\% by \cite{Baines1995} to 4\% by \cite{Rages1991}, and
include latitude dependent values between these values inferred by
KT2009. Further, with the exception of Rages et al., prior modelers
have not followed the occultation constraints on the the vertical
distribution of methane.  This motivates our efforts to redo the
occultation analysis with consideration of a wider range of solutions,
and to examine more carefully the plausibility of a compact methane
cloud layer on Uranus.

In the following, we pursue the point of view that the occultation
measurements of sudden slope changes in refractivity do indicate a
region of sudden changes in methane mixing ratio, which are indicative
of the condensation level of methane, and possibly of a thin region
containing cloud particles.  We describe a reanalysis of the occultation
measurements that can actually achieve saturated vapor pressures in the same region as
the sudden changes in refractivity slope.  We also find 
solutions with high methane amounts at and above the cloud level
that are consistent with the adopted profile of KT2009.
After finding a range of solutions with the desired characteristics, we
then constrain these solutions by calculating spectra for compact
cloud layer models and comparing the pressures inferred from matching
spectra to the pressures inferred from the occultation analysis. We
conclude that a methane cloud layer at the occultation pressure is
consistent with the spectral observations, but that most of the cloud
opacity is concentrated in a deeper layer that was not detected by the
occultation measurements. We also confirm the conclusions of KT2009
that the methane is strongly depleted at high latitudes, but to
shallow depths.

\section{Approach to Occultation Analysis}

We first describe the basic methods of analysis, how we reconstruct
the refractivity profile, then use that profile to validate
our methane retrieval methods by comparison with results of L87.

\subsection{Physical Basis and Methods of Occultation Analysis}

The occultation measurements, after accounting for observing geometry,
can be reduced to refractivity as a function of altitude, where
refractivity is defined to be $N=($index of refraction -1)$\times
10^6$.  If the molecular composition is known, then refractivity can
be converted to number density and mass density.  The pressure can
then be determined by integrating the product of mass density and
gravity, assuming hydrostatic equilibrium.  From pressure and density,
the equation of state of the gas can then be used to infer
temperature.  The main complication is that the variable distribution
of methane is not known and cannot be directly inferred from the
observations. This allows a range of T(P) solutions that depend on
what is assumed about the methane distribution.

The procedure followed by L87 was to first select a
molecular weight that yielded a thermal profile near the tropopause
that was consistent with IRIS thermal infrared observations.  The only
significant constituents at that level of the atmosphere are hydrogen
and helium, so that the molecular weight was determined by their
ratio, which was taken to be the \cite{Conrath1987JGR} value of
He/H$_2$=15/85.  L87 did not consider other He/H$_2$ ratios, even
though the quoted uncertainty is large enough to permit substantially
different profiles, as will be shown in Sec.\ \ref{Sec:vmrvar}.  The
next step in the procedure was to select altitudes bounding the cloud
layer.  These altitudes were not stated by L87, but are
approximately between 5 and 7 km below the 1 bar level determined for
the D model. These are the rough locations of the rapid changes in
refractivity slope.  While the altitude relative to the center of the
planet is fixed for all the profile solutions, the pressure varies
somewhat from one solution to the next, so that the altitude above the
1 bar level also varies slightly.

To constrain the methane profile L87 generally assumed a
constant relative methane humidity above the cloud layer and used the
tropopause mixing ratio to set the constant stratospheric mixing
ratio.  Within the cloud region, L87 state that the
temperature lapse rate was set equal to the wet adiabatic lapse rate,
and adjusted the number density and temperature to match the
refractivity profile.  However, the model D $T(P)$ profile of
L87 does not match the wet adiabat within the assumed
cloud layer.  Instead, the profile matches a weighted average of the
form
\begin{eqnarray}
 (dT/dz)_\mathrm{cld}=(dT/dz)_\mathrm{dry}\times (1-RH) + \\\nonumber
  (dT/dz)_\mathrm{wet}\times RH
\label{Eq:lapse}\end{eqnarray}
where $RH$ is the relative humidity.  This weighted average is
consistent with the suggestion that the occultation sampled a broad
horizontal region in which the average humidity was less than expected
for a uniform cloud layer.  That is also offered as an explanation for
the sub-saturated humidity levels that were obtained in the cloud
layer.  The relatively smooth I/F profiles observed on Uranus
\citep{Sro2007struc, Sro2009eqdyn} are rarely disturbed by discernible
discrete features, especially at low latitudes, and thus this
explanation is not a compelling one.

L87 also introduced a condensed fraction of the methane within the
putative cloud layer, but did not publish the inferred values, nor how
these values were constrained.  If the temperature profile is
constrained to follow the wet adiabat (or the weighted average of wet
and dry adiabats given above) then the only remaining variable that
needs to be adjusted is the fraction of total methane.  There is no
need to partition a fraction of the total into condensed form unless
leaving the total in gaseous form would lead to supersaturation.  In
the latter case, it is reasonable to treat the excess vapor as
condensed material. However, it is not reasonable to allow more than a
tiny fraction of the methane to be in condensed form. The condensed
fractions reaching 10\% or so that are noted by L87 are
grossly inconsistent with near-IR and CCD spectral constraints because
those constraints permit only a low opacity cloud layer.

Even a few percent of methane in condensed form would lead to
extremely large optical depths at the cloud layer, which would
provide very obvious spectral signatures that are not seen. Models of
near-IR and visible spectra require only small optical depths,
typically unity or less at visible wavelengths.
When treated as Mie particles the inferred particle size of a compact
cloud near 1.2 bars is of the order of $r=1$ \mumx.  The total mass
per unit area for a given optical depth $\tau$ is given by
$m=\frac{4}{3}\rho r \tau /Q$, where $\rho$ is the density of solid
methane and $Q$ is the extinction efficiency.
Assuming $r=1$ \mumx, $\tau=1$, $Q=1$, and $\rho= 0.5$ g/cm$^3$
\citep{Costantino1975}, the mass density is 6.7$\times 10^{-5}$
g/cm$^2$. This is a factor of 300,000 times smaller than the typical
20 g/cm$^2$ of total gaseous methane within the cloud layer. Spectral
constraints thus require that the condensed fraction must be so small
as to play an insignificant role in the refractivity profile.

While a $T(P)$ solution consistent with the refractivity measurements
is possible with no methane at all in the atmosphere (Model A of L87),
this was rejected because methane clearly plays a major role in
shaping the spectrum of Uranus.  As the mixing ratio above the cloud
is increased, the inferred temperature must increase to maintain the
observed refractivity, and so does the methane mixing ratio inferred
for the deep atmosphere.  The maximum cloud-top humidity inferred by
L87 is 53\%, which leads to a deep mixing ratio of 4\%, although this
solution did not yield the highest humidity level within the cloud
layer. Trying to increase the above-cloud humidity any further results
in rapidly increasing temperatures into the cloud layer and
unacceptable superadiabatic lapse rates. None of these profiles yield
anything close to saturation at the altitudes where cloud condensation
is suspected.

\subsection{Reconstructing the refractivity profile.}

\begin{figure}\centering
\includegraphics[width=3.2in]{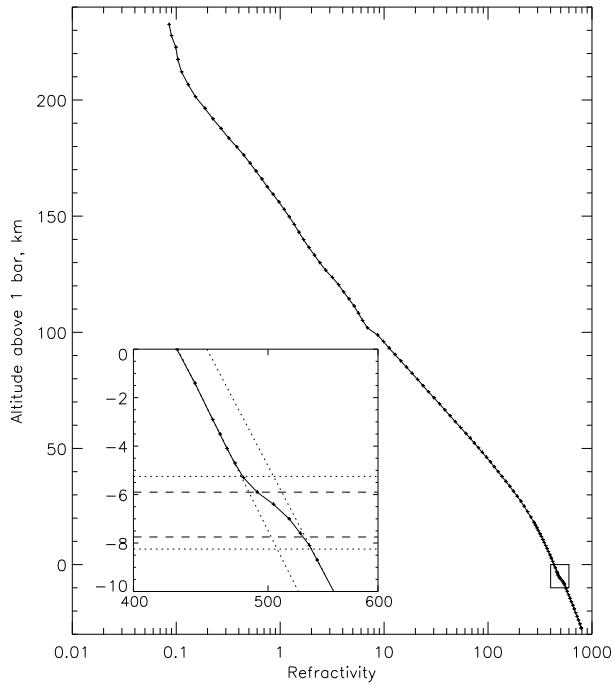}
\caption{Refractivity profile computed from the L87 model D
profile. The inset provides a detailed view of the sudden slope
change, and there horizontal lines indicate cloud boundaries for our
D1 (dotted) and F1 (dashed) structure solutions (discussed in Sec. 3).}
\label{Fig:refracprof}
\end{figure}

With the aim of conducting a reanalysis of the occultation profile
with different assumptions, we first needed to create a detailed
refractivity profile.  We began with the tabulation of P, T, molecular
weight, number density, and mixing ratio published by L87.
We used refractivity values per molecule of $K_{He}$= 0.5062,
$K_{H_2}$ =0.1302, and $K_{CH_4}$=1.629 (all in units of 10$^{-17}$
cm$^{-3}$), which are the same as those used by L87 and
referenced therein.
We then used the relation
\begin{eqnarray} K(z) = K_{H_2} f_{H_2}(z) + K_{He} f_{He}(z) +
K_{CH_4} f_{CH_4}(z)\label{Eq:refrac}
\end{eqnarray}
to compute the mean molecular refractivity, where subscripted $f$ values denote
numeric fractions for each molecule. We also used the same 15/85 ratio of He to H$_2$
as L87.  The total refractivity for the mixture
is then given by \begin{eqnarray}
   N(z) = n(z) K(z),
\end{eqnarray}
where $n(z)$ is the total number density at altitude $z$. Our computed refractivity profile versus
altitude is shown in Fig.\ \ref{Fig:refracprof}.

\begin{figure}\centering
\includegraphics[width=3.2in]{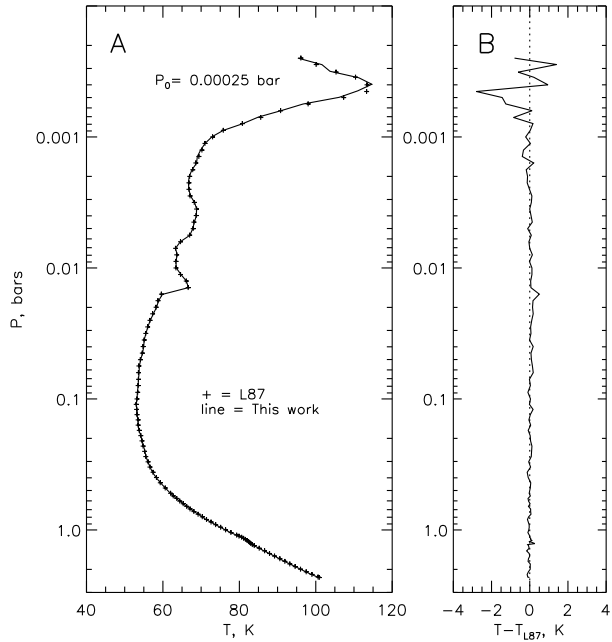}
\caption{A: $T(P)$ profile derived from refractivity profile of
Fig.\ \ref{Fig:refracprof} (solid) using L87 tabulated
results, and the L87 model D profile (symbols). B: the
difference profile.}
\label{Fig:tpvalidation}
\end{figure}

To validate that the refractivity profile we computed was consistent with the temperature
profile, we inverted the profile as follows.  We computed the pressure vs altitude under
the assumption of hydrostatic equilibrium using \begin{eqnarray}
  P(z) = P_0+\int_{z_0}^z M(z)(n(z)/N_A)g(z)dz ,\label{Eq:pint}
\end{eqnarray} 
where M(z) is the molecular weight in grams per mole, $N_A$ is
Avogadro's number, and $g(z)$ is the gravitational acceleration as a
function of altitude, which varies from 8.6843 m/s$^2$ at 1 bar (0 km)
to 8.5157 m/s$^2$ at 240 km, assuming a retrograde wind of 100 m/s
\citep{Sro2009eqdyn}, which reduces $g$ by only 0.23\% and could be
ignored.  We started the downward integration at 240 km above the
1-bar level, and took the starting pressure to be 2.5$\times10^{-4}$ bar to match
the stratospheric profile of L87. (The specific value of $P_0$ has an
insignificant effect on the structure for pressures greater than a few
hundred millibars.) Assuming an ideal gas equation of state we then computed
$T(P)$ from pressure and number density. Our $T(P)$ profile thus
constructed is compared with the L87 profile in Fig.\
\ref{Fig:tpvalidation}, where we see that differences below the
tropopause are generally smaller than 0.1\deg (the RMS deviation is
0.07 K).  This provides a reasonable validation of our reconstructed
refractivity profile, which we will reanalyze with different
assumptions in a subsequent section, after first validating our
methane retrieval procedure.

\subsection{Validation of methane profile retrieval}

We next tried to reproduce the methane profile retrievals obtained by
L87.  We started with the refractivity profile, the
assumed He/H$_2$ ratio of 15/85, selected altitudes for the cloudy
layer, and selected a constant relative humidity above the cloud layer
up to the tropopause, and above the tropopause assumed a constant
methane mixing ratio equal to the tropopause value, although later we
used the 10$^{-5}$ upper limit of \cite{Orton1987}.  Within the cloud layer we
forced the lapse rate to agree with Eq.\ \ref{Eq:lapse}.  We then
started at the top of the atmosphere and integrated downward the
number density and pressure and iteratively solving for $f_{CH_4}$ and
$n(z)$ under the constraints of the specified methane humidity above
the cloud, the constraints of the temperature lapse rate within the
cloud, and the fixed mixing ratio below the cloud, which was taken to
be the mixing ratio at the bottom of the cloud layer.  

\begin{figure*}\centering
\includegraphics[width=5in]{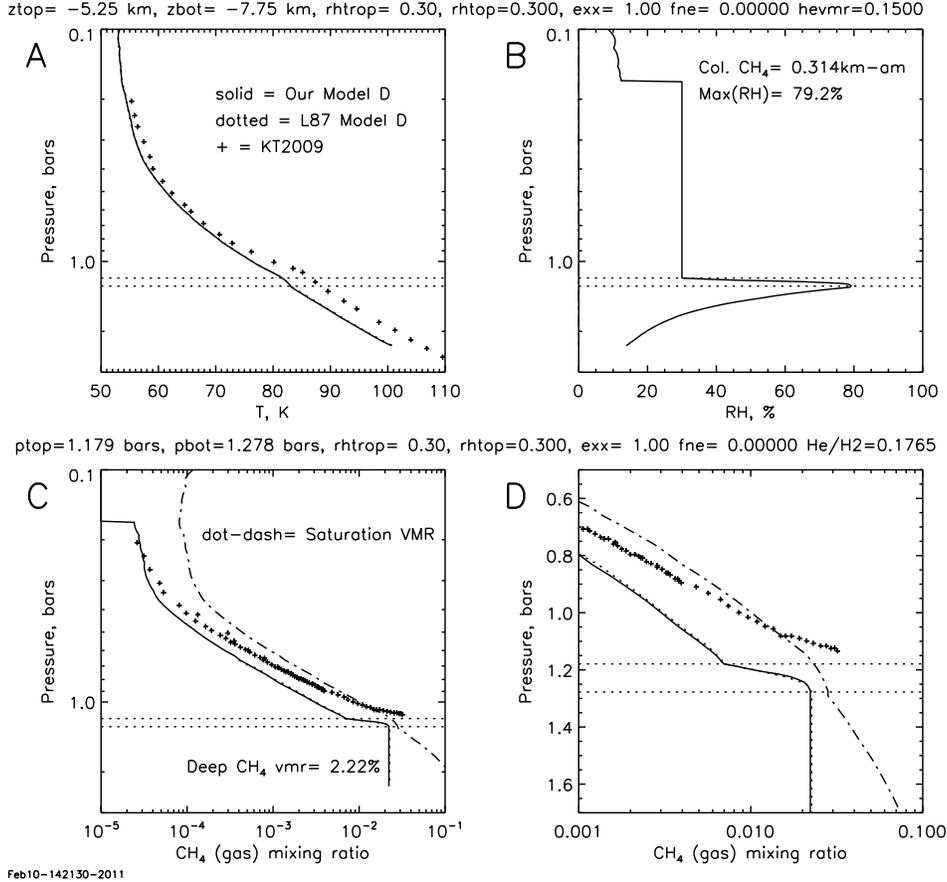}
\caption{A: Inverted $T(P)$ profile for 30\% RH above cloud (our Model
D) compared to the L87 Model D (dotted) and the profile adopted
by KT2009 (+). B: relative humidity profile of
the inverted $T(P)$ profile. C: Methane mixing ratio profile we
inverted (solid) compared to the L87 Model D (dotted) and that
adopted by KT2009 (+), with the saturation mixing
ratio shown as the dot-dash curve, using our $T(P)$ profile. D:
 detailed views of the methane mixing ratio profiles. The
extra dashed curve is the saturation methane mixing ratio computed
from the KT2009 $T(P)$ profile.  Note the very close
agreement between our inversion and that of L87, in
several cases too close to distinguish their difference. The
horizontal dotted lines are at pressures of 1.179 bars and 1.278 bars,
which correspond to altitudes of 5.25 km and 7.75 km below the 1-bar
level.}
\label{Fig:Dinversion}
\end{figure*}

Our attempt to reproduce the Model D solution is displayed in Fig.\
\ref{Fig:Dinversion}.  In most respects our results are barely
distinguishable from those of L87.  We obtain a deep
\chf VMR of 2.22\% compared to their value of 2.26\%, and our maximum
\chf RH at the cloud base is 79.2\% compared to their 78\%.  These might be
brought into better agreement with slightly different choices for the
cloud boundaries. We did not assign any of the methane fraction to
condensed material; we don't know whether L87 did or not for
this model. We take these comparisons to be adequate validation of our
inversion technique.  Note that the $T(P)$ and \chf profiles of
KT2009, which are also plotted in Fig.\
\ref{Fig:Dinversion}, have significantly higher temperatures and much
more methane above the cloud top.

If we increase the above cloud relative humidity as much as plausible,
which we estimate to be about 57\% (instead of the 53\% of
L87), we get results very close to the L87 Model F
profile. We obtained a deep \chf VMR of 4.13\% (instead of 4\%) and a
peak in-cloud humidity of 70\% instead of 72\%. The temperature
profile is also close to the Model F profile of L87.
Even at this upper limit, however, the above cloud methane is well
below what was adopted by KT2009. Pushing the
methane values above this level results in physically unacceptable
results: the in-cloud humidity becomes lower than the humidity above
the cloud and the sub-cloud lapse rate becomes more and more unstable.
We would argue that even this upper limit leads to physically
implausible profiles because of the relatively low humidity in the
cloud layer.  Yet good fits to the spectral observations seem to need
even more methane at these levels.  In the following section we show
how these levels can be reached.

\section{Revised analysis of the refractivity profile.}\label{Sec:vmrvar}

Now that we have validated our analysis techniques, we apply these
techniques to obtain new solutions for temperature and methane
profiles.  We first revised the methane mixing ratio at the tropopause
and throughout the stratosphere to equal the \cite{Orton1987} upper
limit of 1$\times 10^{-5}$, and used a variable relative humidity
between the tropopause and the cloud top, using the formulation
\begin{eqnarray}
RH(z) = RH_{trop}+(RH_{ctop}-RH_{trop})(1-\\\nonumber
(z-z_{ctop})/(z_{trop}-z_{ctop}))^x,
\end{eqnarray} 
where a tropopause humidity $RH_{trop}$ of about 12\% is needed to
match the desired minimum methane VMR, and the tropopause height
$z_{ctop}$ is taken to be 45 km. An exponent $x$=1 provides linear
interpolation, and a decrease of humidity above the cloud top that is
similar to that adopted by KT2009.  These changes
made no significant difference in the plausible upper limit of methane
mixing ratios in the vicinity of the cloud layer and in the deep
atmosphere.  We also tried adding neon to the atmosphere, using the
mixing ratio of 0.0004 suggested by \cite{Conrath1987JGR}. This has a
very small but undesirable effect that reduces the upper limit on
methane.  What is really needed is a lower background molecular
weight, which then requires increased methane to match the
refractivity profile.  The only plausible way to obtain a lower
molecular weight is to decrease the mixing ratio of He, which is
what we proceeded to do in the following fashion.  Within the cloud
layer we used the same formulation as given by Eq.\ \ref{Eq:lapse},
but decreased the He VMR (at all altitudes) as needed to obtain methane
condensation within most of the putative cloud layer.  Below the cloud
layer we used the same methane mixing ratio as found at the bottom of
the cloud layer.

\begin{figure*}\centering
\includegraphics[width=5in]{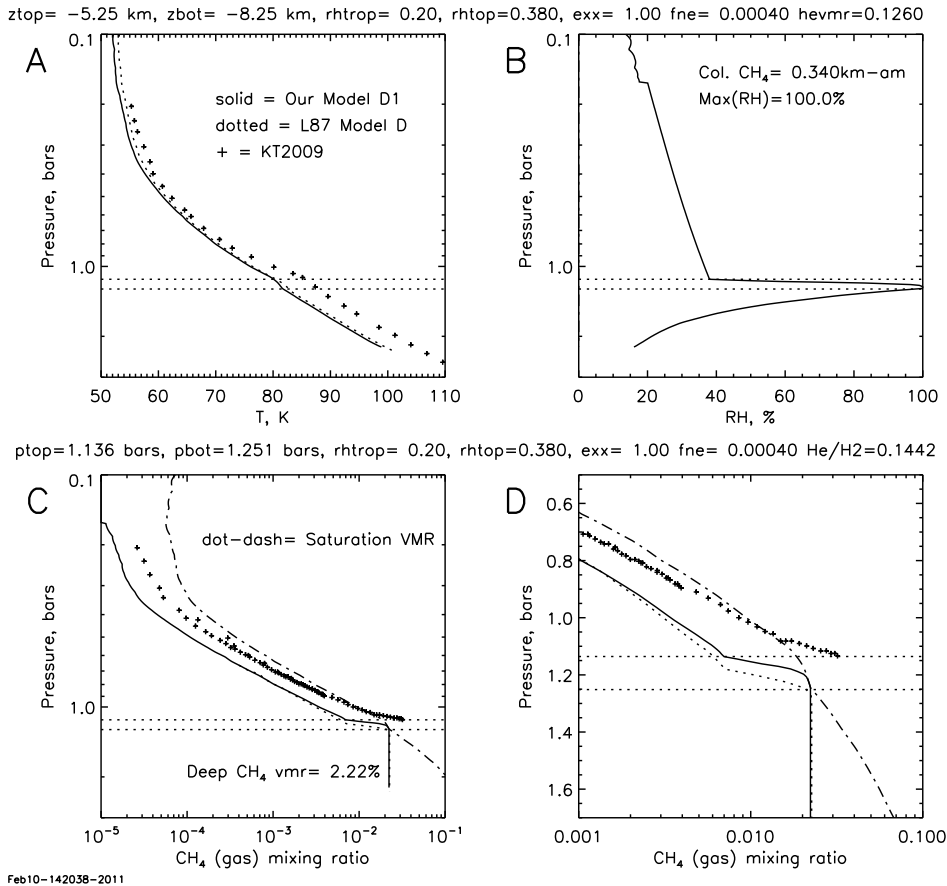}
\caption{As in Fig.\ \ref{Fig:Dinversion} except that we plot our
Model D1 profile, which uses a volume mixing ratio of 0.126 for He, an
above cloud humidity of 38\%, linearly interpolated to a tropopause
humidity of 12\%.  The horizontal dotted lines are at pressures of
1.136 bars and 1.251 bars, which correspond to altitudes of 5.25 and
8.25 km below the 1-bar level.}
\label{Fig:ModelD2}
\end{figure*}

Decreasing the He mixing ratio has two beneficial effects: it allows methane
saturation mixing ratios to be attained within the layer where
condensation is expected to occur, and it allows a higher maximum
methane mixing ratio that is more likely to be consistent with near-IR and
visible spectra of Uranus.  Our first example solution (Model D1, Fig.\ \ref{Fig:ModelD2}) uses
a He VMR of 0.126, and yields a saturated methane mixing ratio through the bottom half of
the nominal cloud region, and a deep mixing ratio of 2.22\% for an above-cloud
humidity of 38\%.  This is similar to Model D of L87, but is more
physically plausible.

\begin{figure*}\centering
\includegraphics[width=5in]{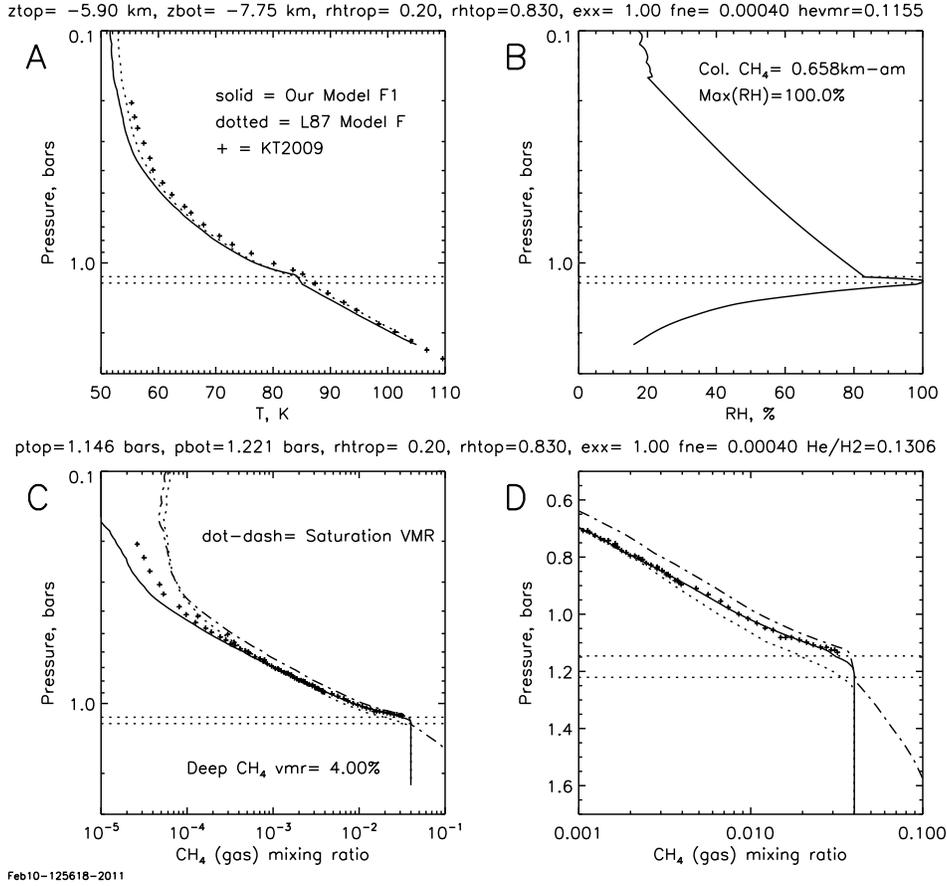}
\caption{As in Fig.\ \ref{Fig:ModelD2} except that we show our Model
F1 profile, which uses a volume mixing ratio of 0.1155 for He, and an above
cloud humidity of 83\%.  The horizontal dotted lines are at pressures
of 1.179 bars and 1.278 bars, which correspond to altitudes of 5.25
and 7.75 km below the 1-bar level.}
\label{Fig:ModelF2}
\end{figure*}

A second example (Model F1), which attains the same deep mixing ratio of Model F
(4\%) is shown in Fig.\ \ref{Fig:ModelF2}.  This solution is notable
in having much more methane above the cloud layer than Model F, and
provides a close match to the \chf VMR profile adopted by
KT2009, although our corresponding $T(P)$ profile
is somewhat cooler than their adopted profile, as needed to match the
refractivity profile.  The He VMR for this solution is 0.1155, which
is only 1.05$\sigma$ below the nominal value of 0.15$\pm$0.033, given
by \cite{Conrath1987JGR}. Our maximum methane solution (our Model G in
Table\ \ref{Tbl:strucsum}) 
is obtained for a He VMR of 0.1063, which is only
1.3$\sigma$ below the nominal value.  This solution has an above-cloud humidity of
100\%.  This provides a deep \chf VMR of 4.88\% and even more methane
above the cloud top than that adopted by KT2009.

\begin{figure*}\centering
\includegraphics[width=5.5in]{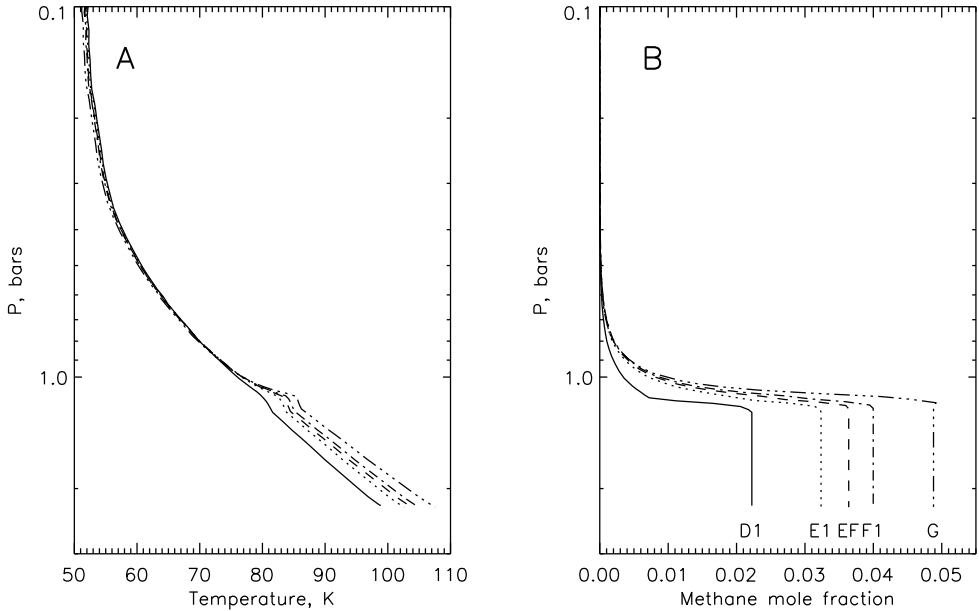}
\caption{A summary plot comparing the profiles of temperature (left) and methane
mixing ratios (right) for models D1, E1, EF, F1, and G.}
\label{Fig:profcomp}
\end{figure*}

A summary of the above model profiles and intermediate model EF is
provided in Fig.\ \ref{Fig:profcomp} and detailed parameter
information in Table\ \ref{Tbl:strucsum}. The small difference in
tropopause temperatures (at $p$ = 0.1 bar) is due to the different
helium mixing ratios used in each model.  The maximum temperatures
reached near $p$= 2.3 bars for our models D1-G are comparable to those
obtained by L87 for their models C-F.

We now have a range of solutions that are consistent with occultation results,
as well as being consistent with the expectation that methane humidity levels should
reach saturation levels in the putative condensation region, which is presumably
the physical reason for the sudden change in refractivity slope.  The
remaining question is whether it is possible to fit the spectra of Uranus with a cloud
particle layer in the same pressure regime where occultation analysis implies a cloud
layer. We first describe our chosen spectral constraints and radiation transfer
and fitting methods, and then proceed to describe the results of applying the
spectral constraints. 

\section{Use of Uranus spectral constraints}

\subsection{Spectral observations}
\begin{figure*}\centering
\includegraphics[width=5in]{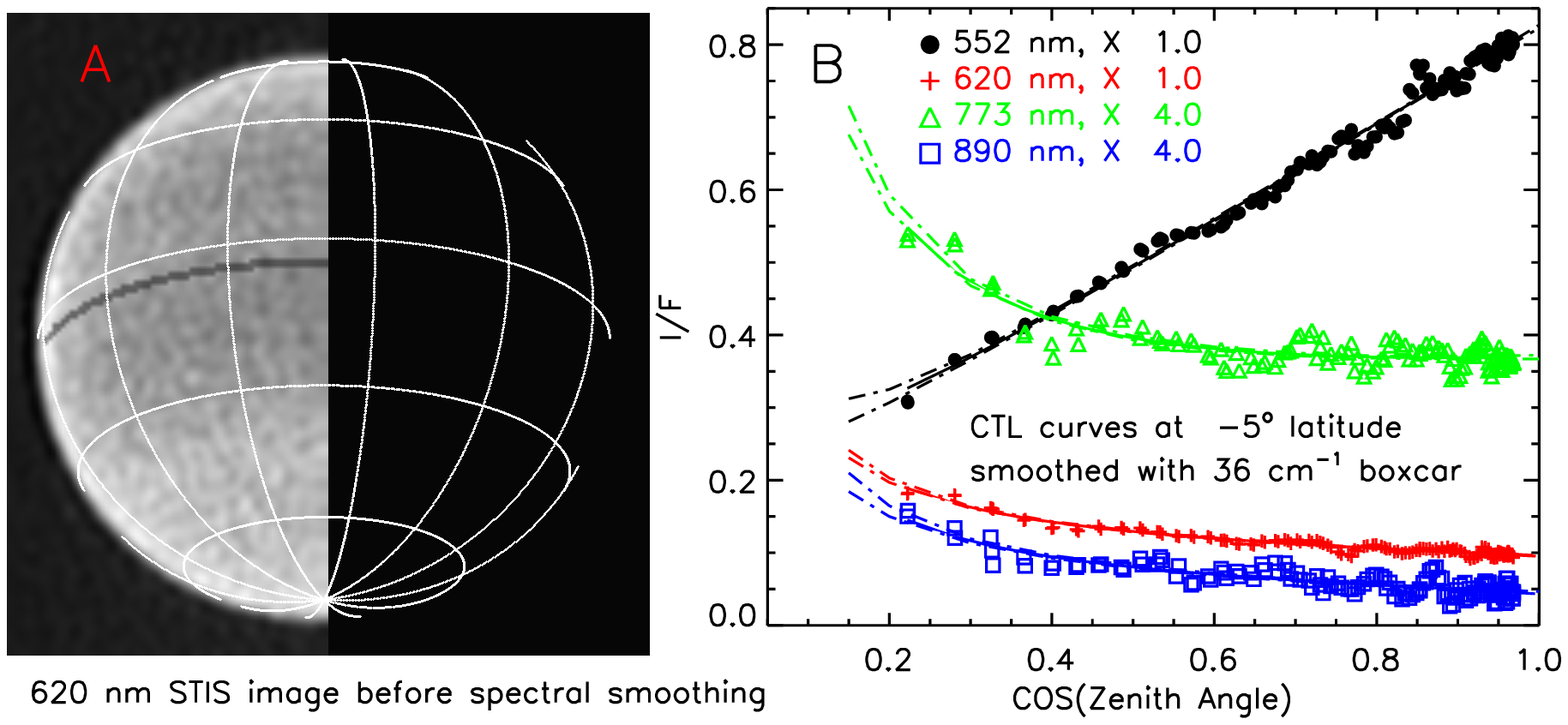}\\
\hspace{-0.5in}\includegraphics[width=5in]{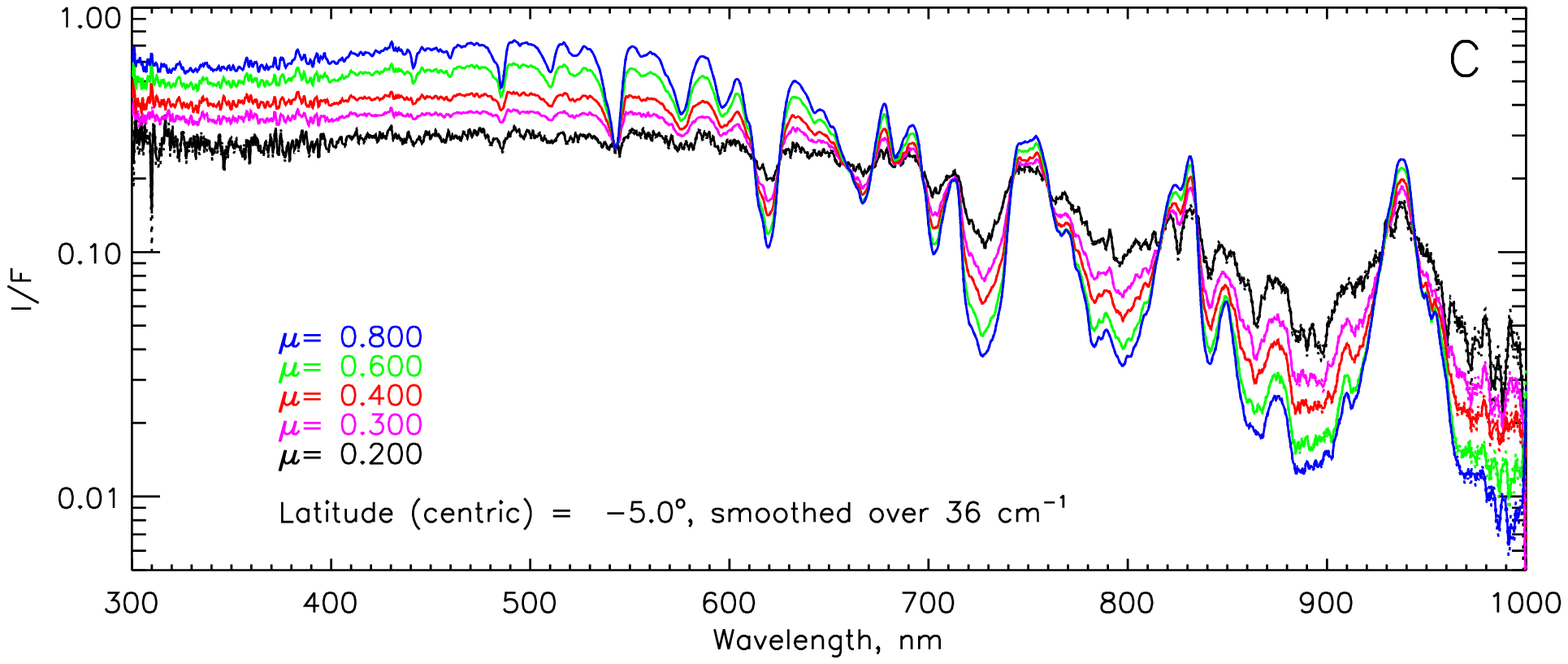}
\caption{A: Sample image from the unsmoothed KT2009 STIS data cube at
620 nm, with grayed pixels indicating our center-to-limb sampling at
5\deg S, with grid lines at 30\deg intervals. B: Sample center-to-limb
scans at four wavelengths (after spectral smoothing with a 36 \icm
boxcar), with fits shown by dot-dash lines bounding 1$\sigma$
uncertainty limits. C: Sample interpolated STIS I/F spectra at 5\deg S planetocentric
latitude after CTL fitting and spectral smoothing.}
\label{Fig:specobs}
\end{figure*}

We chose for our spectral comparisons the spatially resolved STIS
observations made on 19 August 2002, as corrected and calibrated by
KT2009.  These spectra are undoubtedly the most accurately calibrated
and characterized spectra available for the purpose.
The data provide two spatial dimensions, with a pixel size of 0.05 arc
seconds on a side, providing 37 samples from center to limb,
and one spectral dimension, which is sampled at 0.4-nm intervals,
providing 1-nm resolution from 300 nm to 1000 nm. A sample image from
this cube is provided in Fig.\ \ref{Fig:specobs}A.
These data provide a good view of the low latitude region sampled by
the Voyager radio occultation experiment, although some 16 years
later, which is a delay of 20\% of a Uranus year. However, given the
long radiative time constants in the Uranus atmosphere
\citep{Conrath1990}, it is plausible to expect only small changes in
cloud structure in low latitude regions. Stability at low latitudes
also seems to be compatible with the analysis of HST images from 1994
to 2000 \cite{Kark2001IcarUrSeas}, which indicate little change in
albedo structure at these latitudes.  

In addition to spectral constraints, the KT2009 data provide important
center-to-limb (CTL) information (Fig.\ \ref{Fig:specobs}B). To reduce
the effects of noise, we fit the center-to-limb scans to a smooth
function of $\mu$ (the cosine of the zenith angle), as illustrated in
Fig.\ \ref{Fig:specobs}B for a planetocentric latitude of 5\deg S. We
then sampled that function at $\mu$ = 0.3, 0.4, 0.6, and 0.8.  The
resulting substantial noise reduction is readily apparent in the
figure, where uncertainties are shown by roughly parallel dot-dash
lines at 1$\sigma$ limits.  We were able to ignore the difference
between solar and observer zenith angles because the STIS observations
were taken at a very small phase angle (0.04\degx).  We also were able
to ignore possible longitudinal variations in cloud structure because
the discrete cloud features of Uranus are generally so small and of
such low contrast that they do not obscure the CTL information of the
zonal bands. We also know from prior experience with differences
between Uranus images taken at substantially different central
meridian longitudes that the only substantial longitudinal I/F
variations are due to discrete features. At the few latitudes where
discrete features were found in the STIS data the fits easily
interpolated across them.  The selected 5\deg S STIS spectra
interpolated to five cosine values are displayed in Fig.\
\ref{Fig:specobs}.  Note the strong limb darkening at short
wavelengths and the strong limb brightening at centers of the methane
absorption bands at longer wavelengths. The spectrum for $\mu$ = 0.2,
which is noisier and not well corrected because it is too close to the
limb, was not used in our analysis.

\begin{table*}\centering
\caption{Occultation analysis parameters and characteristic results for our models.}
\vspace{0.15in}
\begin{tabular}{r c c c c c c}
\hline\\[-0.1in]
Model name:            &  D    &   D1  &  E1   &  EF   &  F1 & G \\[0.05in]
         Tropopause RH & 30\%  & 20\%  & 20\%  & 20\%  & 20\%  & 20\%  \\[0.05in] 
          Cloud top RH & 30\%  & 38\%  & 70\%  & 78\%  & 83\%  &100\%  \\[0.05in] 
            Maximum RH & 79\%  & 100\% & 100\% & 100\% & 100\% & 100\%\\[0.05in]
              Neon VMR & 0.0   &0.0004 & 0.0004& 0.0004& 0.0004& 0.0004 \\[0.05in]
          Cloud Top,km & -5.25 &-5.25  &-5.50  &-5.70  &-5.90  &-6.50  \\[0.05in]
         Cloud Bot.,km & -7.75 &-8.25  &-8.00  &-7.80  &-7.75  &-7.75  \\[0.05in]
        Cloud Top, bar &1.179  &1.136  &1.142  &1.142  &1.146  &1.153  \\[0.05in]
       Cloud Bot., bar &1.278  &1.251  &1.241  &1.226  &1.221  &1.205 \\[0.05in]
                He VMR & 0.15  &0.126  &0.122  &0.1179 &0.1155 &0.1063 \\[0.05in]
 $\Delta \mathrm{VMR}/\sigma_\mathrm{VMR}$ 
                      & 0.00 & -0.73 & -0.85 & -0.97 & -1.05 & -1.32\\[0.05in]
             He/H$_2$ & 0.1765&0.1442 &0.1390 &0.1337 &0.1306 &0.1189\\[0.05in]
         Deep \chf VMR &2.22\% &2.22\% &3.24\% &3.64\% &4.00\% &4.88\% \\[0.05in]
\chf km-am to Cld Bot. & 0.314 & 0.340 & 0.576 & 0.610 & 0.658 & 0.719\\[0.05in]
\hline\\[-0.1in]
\end{tabular}\label{Tbl:strucsum}
\end{table*}

To further reduce noise in the observations, and to facilitate model
comparisons, we also smoothed the observed spectra to a uniform
wavenumber resolution using a 36 \icm boxcar. A uniform wavenumber
resolution was chosen to fit the requirements of our Raman scattering
code \citep{Sro2005raman}, and the specific resolution is chosen to be both commensurate
with the wavenumber shifts of the three most important Raman
transitions and to approximately match the resolution provided by the
STIS observations at 550 nm.  The spectral smoothing and sampling of
the fits to the CTL variations made noise in the observations a fairly
small contributor to the total uncertainty as demonstrated in Sec. 4.4.

\begin{figure}\centering
\includegraphics[width=3.4in]{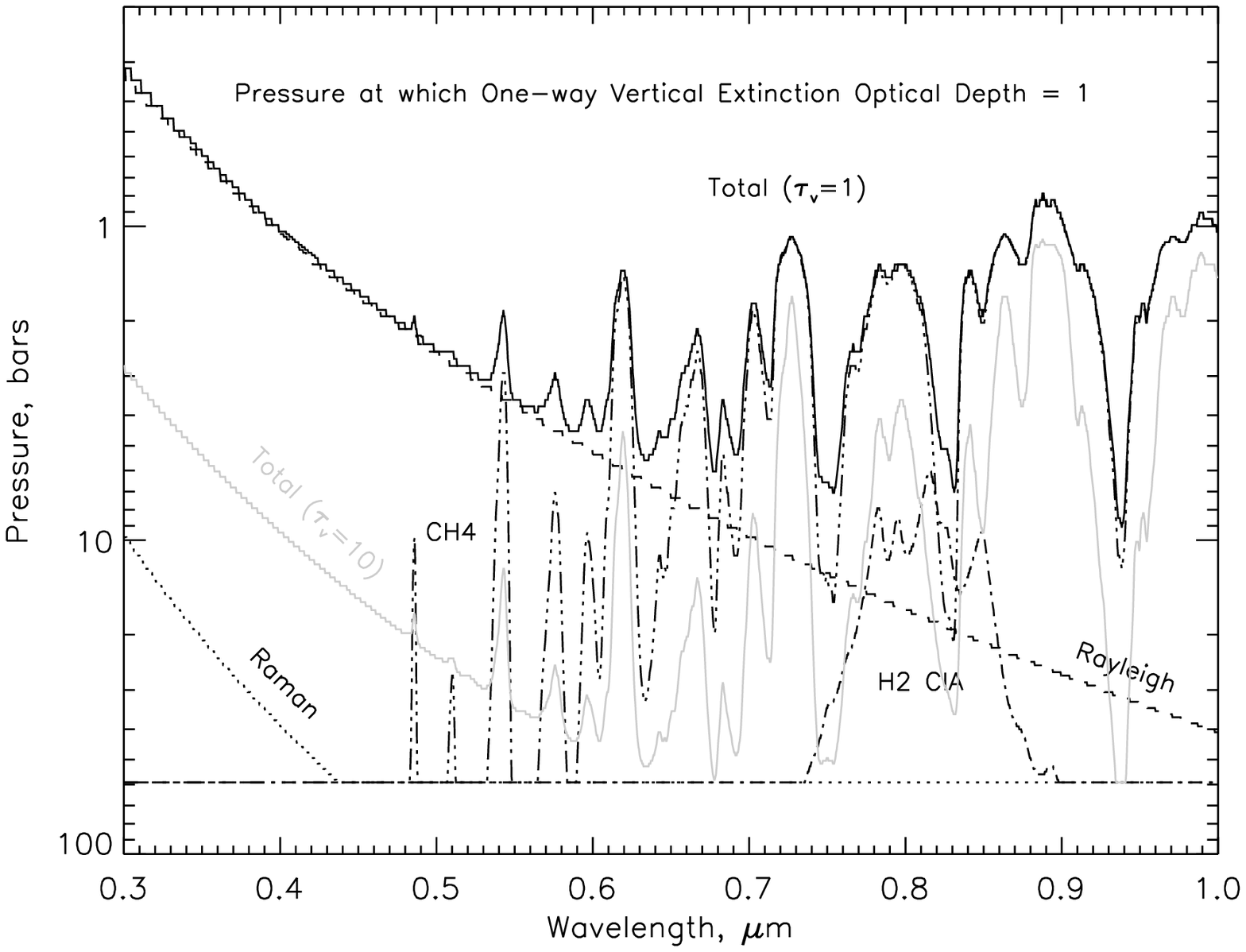}
\caption{Pressure at which the vertical optical depth to space
reached unity, shown for individual components including Raman scattering (dotted),
Rayleigh scattering (dashed), methane absorption (triple-dot-dash), and H$_2$ CIA (dot-dash).
The pressure for  all effects combined is shown by the dark solid curve for unit
optical depth and by the light solid curve for $\tau=10$.}
\label{Fig:pendepth}
\end{figure}

The sensitivity of these spectra to different atmospheric levels on
Uranus is indicated in Fig.\ \ref{Fig:pendepth}.  This indicates the
penetration depth of light into an aerosol-free atmosphere of Uranus
by the plot of pressures at which one-way unit vertical optical depth
is reached. Penetration depths for individual contributions by
Rayleigh scattering, methane absorption and hydrogen collision-induced
absorption (H$_2$ CIA) are also shown.  A key wavelength region is
near 0.825 \mumx, where H$_2$ CIA opacity exceeds the opacity of
methane.  Poor fits in this region are an indication of incompatible
vertical distributions of methane and hydrogen absorptions,
and may be a result of assuming an incorrect methane mixing ratio.

\subsection{Radiation transfer calculations}
We used the radiation transfer code described by \cite{Sro2005raman},
which include Raman scattering and polarization effects on outgoing
intensity.  To save computational time we employed the accurate
polarization correction described by \cite{Sro2005pol}.  After trial
calculations to determine the effect of different quadrature schemes
on the computed spectra, we decided to use 14 zenith angle quadrature
points per hemisphere and a 14-order azimuthal expansion for our
compact model and 10 quadrature points for fitting models with the
KT2009 structure.  To characterize methane absorption we used the
corrected coefficients of KT2009.  To model collision-induced
absorption (CIA) of H$_2$-H$_2$ and He-H$_2$ interactions we
interpolated tables of absorption coefficients as a function of
pressure and temperature that were computed with a program provided by
Alexandra Borysow \citep{Borysow2000}, and available at the
Atmospheres Node of NASA'S Planetary Data System. We assumed
equilibrium hydrogen for most calculations but did look into the
effects of non-equilibrium distributions, which are discussed in
Sec. \ref{Sec:noneqh2}

\subsection{Cloud models}

We used two distinct models of cloud structure for comparison
purposes. The first is nearly identical to the model of
KT2009, which we will refer to as the KT2009 model, and the second
is a modified version, which we will refer to as the compact cloud layer
model. A plot of the vertical distribution of these layers can be
found in Sec. 5.3 (Fig. 14).

\subsubsection{The KT2009 model}
\begin{figure}\centering
\includegraphics[width=3.2in]{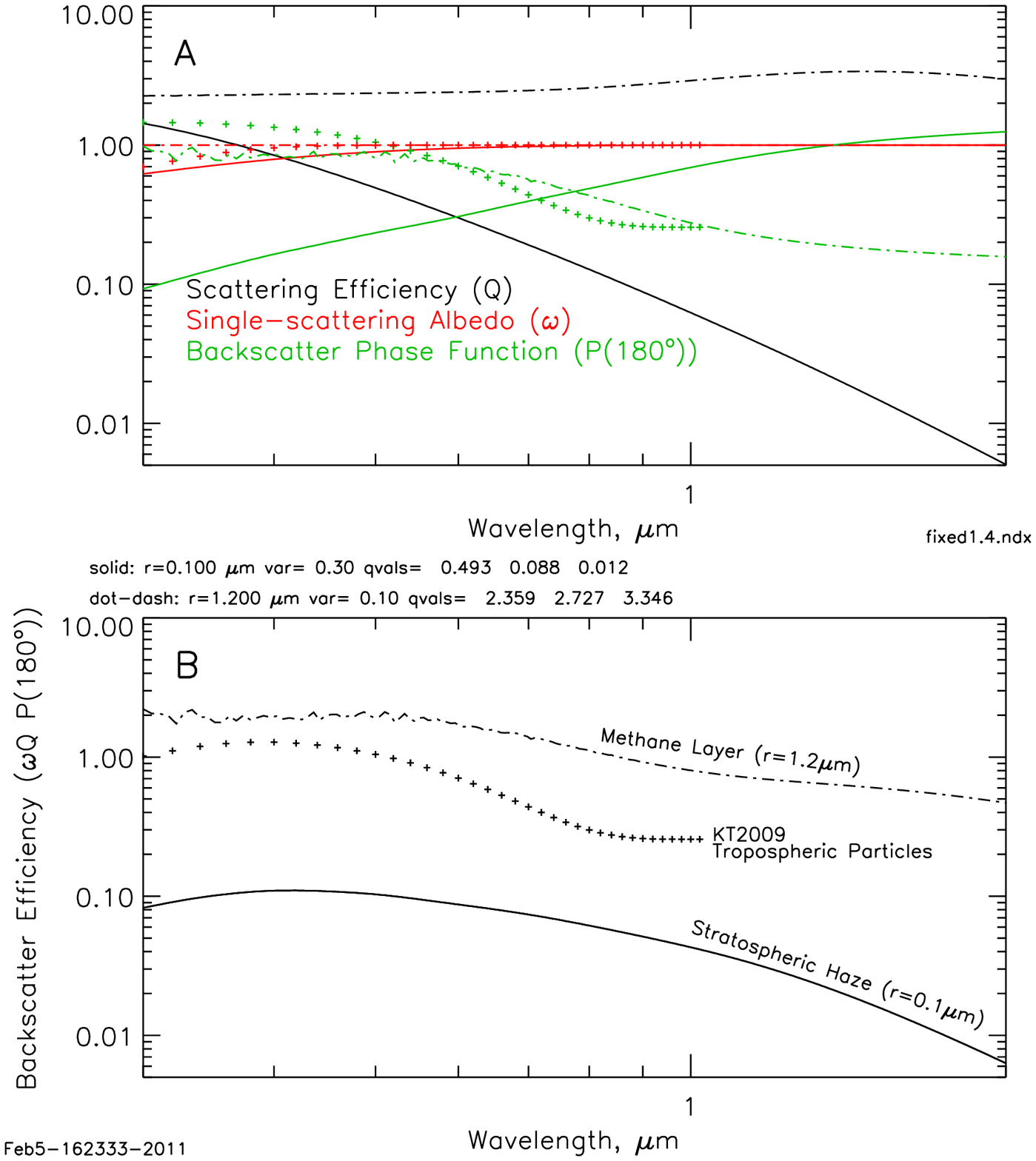}
\caption{Scattering properties of the stratospheric haze model
(solid), the middle tropospheric Mie layer (dot-dash) and KT2009
tropospheric particles (plus signs). A: extinction efficiency (Q,
black), single-scattering albedo ($\varpi$, red) and backscatter phase
function ($P(\pi)$, green). B: backscatter efficiency.}
\label{Fig:scatparams}
\end{figure}

This model has four layers of aerosols, the uppermost being a
Mie-scattering stratospheric haze layer characterized by an optical
depth at 0.9 \mumx, a gamma size distribution \citep{Hansen1971JAScircpol}, with a mean
radius of $a=$0.1 \mum and a normalized variance of $b=$0.3. These
particles are assumed to have a real
index of 1.4, and an imaginary index following the KT2009 relation
\begin{eqnarray}
n_i(\lambda)=0.055\exp[(350-\lambda)/100]\label{Eq:ni},
\end{eqnarray}
for $\lambda$ in nm. This haze was distributed vertically above the
100 mb level with a constant optical depth per bar.  The remaining
layers in the KT2009 model are characterized by a wavelength-independent
optical depth per bar and a wavelength-dependent single-scattering albedo,
given by
\begin{eqnarray}
\varpi_t(\lambda)=1-1/[2+\exp[(\lambda-290)/37]],\label{Eq:ssa}
\end{eqnarray} 
again for $\lambda$ in nm.  Their adopted double Henyey-Greenstein
 phase function for the tropospheric layers used $g_1=$0.7,
 $g_2$=-0.3, and a wavelength-dependent fraction for the first term,
 given by
\begin{eqnarray}
f_1(\lambda)=0.94-0.47 \sin^4[(1000-\lambda)/445],\label{Eq:f1}
\end{eqnarray} 
which produces a backscatter that decreases with wavelength, as shown
in Fig.\ \ref{Fig:scatparams}. The three
tropospheric layers are uniformly mixed with gas molecules, with
different optical depths per bar in three distinct layers: 0.1-1.2
bars (upper troposphere), 1.2-2 bars (middle troposphere), and P$>$2
bars (lower troposphere).  These optical depths are the adjustable
parameters we use to fit this model to the observations.

\subsubsection{The compact cloud layer model}

This model is a modification of the KT2009 model.  The main change we
made is to replace their middle tropospheric layer with two compact
layers: an upper tropospheric compact cloud layer and a middle
tropospheric compact cloud layer. This allows the possibility of a
better match to the observations if the aerosols between 1 and 2 bars
do not match the KT2009 assumption of being uniformly mixed with the
gas.  The upper tropospheric cloud (UTC) in our model is composed of
Mie particles, which we characterized by a gamma size distribution
with a mean particle radius of 1.2 \mum and a fixed normalized
variance of 0.1, a fixed refractive index of 1.4, and an imaginary
index of zero. The particle radius was fixed at the given value
because it did not vary much from that value in preliminary fits.  For
the middle tropospheric cloud (MTC) in our model we used particles
with the same scattering properties as given by KT2009 for their
tropospheric particles.  Both of these compact layers have the bottom
pressure as a free (adjustable) parameter and a top pressure that is a
fixed fraction of 0.93 of the bottom pressure.  This degree of
confinement is approximately the same as obtained for the cloud layer
inferred from the occultation analysis. We assumed similar confinement
for the deeper layer in our model, but it could easily be more
vertically diffuse than we assumed, as long as the effective pressure
is similar.

The last change we made was to replace their lower tropospheric layer
by a compact cloud layer at 5 bars (the LTC), with adjustable optical
depth and with the KT2009 tropospheric scattering properties. We found
that this layer was needed to provide accurate fits near 0.56 and 0.59
\mumx, but its pressure is not well constrained by the observations
(the effect of varying the pressure can be compensated by varying the
optical depth, to produce essentially the same fit quality). Whether
this deep cloud is vertically diffuse or compact also cannot be well
constrained, and thus our assumption of a compact cloud for this layer
is a matter of convenience and is not compelled by observations.  The
wavelength dependence of the backscatter phase function, extinction
efficiency, and single scattering albedo of the stratospheric haze are
given in Fig.\ \ref{Fig:scatparams} for both the Mie particles we used
for the putative methane layer (the UTC) and for the KT2009
tropospheric particles.  The latter have wavelength independent
optical depth, and thus the way its backscatter efficiency varies with
wavelength is entirely determined by the phase function (defined by
Eq.\ \ref{Eq:f1}). The best-fit Mie particle size results in a smaller
variation in backscatter efficiency, although both decrease with
wavelength.

In summary, we consider two models.  The diffuse one has the KT2009
structure, which provides a fitting standard of comparison.  The
compact model, the main feature of which is the splitting of the
middle tropospheric layer of KT2009 into two layers, allows us to see
if a compact layer of methane particles can provide good fits to the
observed spectra, and to see which occultation-derived profile of
temperature and methane mixing ratio provides (1) the best fit to the
spectra and (2) the best agreement between the fit pressure for the
middle tropospheric layer and the pressure inferred from the
occultation analysis.  Hopefully, the best spectral fit would occur
for the same profile that provided the best pressure match.  As shown
in the following, that is roughly what happened.

\subsection{Error model}

\begin{figure}\centering
\includegraphics[width=3.2in]{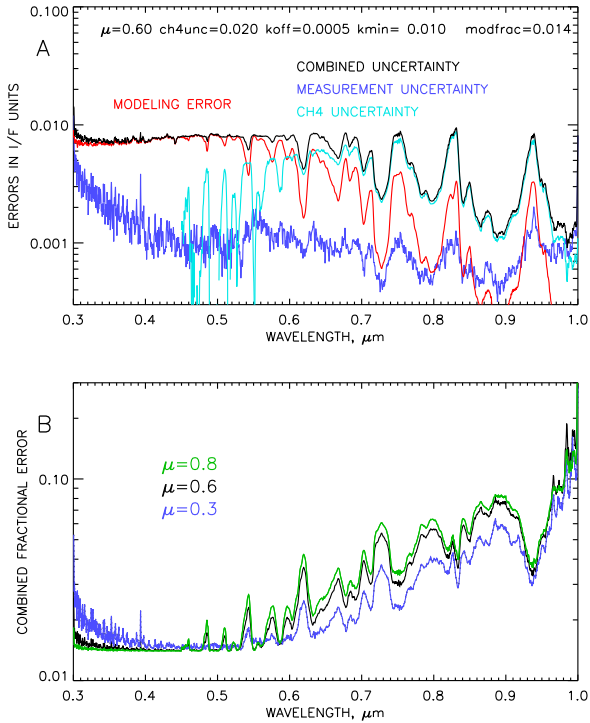}
\caption{A: Estimated contributions by measurement
uncertainty (purple), methane coefficient uncertainty (blue), and overall
modeling errors (red), to the combined relative I/F uncertainty for $\mu$=0.6 (black). B:
Combined Fractional error vs. wavelength at zenith angle cosines of 0.3 (purple),
0.6 (black), and 0.8 (green).}
\label{Fig:comberr}
\end{figure}

To measure fit quality we use \chisqx, which requires an estimate of
the expected difference between a model and the observations due to
the uncertainties in both. We roughly characterized the uncertainty in
measurements by the expected uncertainty in the samples of the smooth
fits, as described previously.  This uncertainty is shown by the
purple curve in the upper panel of Fig.\ \ref{Fig:comberr}. There is
also an overall calibration uncertainty, estimated by KT2009 to be
5\%. However, calibration errors for spectra are similar to nearly
wavelength-independent scale factor errors, which tend to cause
changes in inferred optical depths of aerosols, with very little
effect on their inferred vertical distributions, and thus we did not
include this as an error source. KT2009 estimated relative
uncertainties of their corrected spectra to be within 1\%, which were
estimated by comparison of the corrected spectrally weighted STIS
observations to band-pass filtered images. To this relative
calibration uncertainty we added an overall modeling uncertainty of
1\% for a combined relative fractional error of 1.4\%, shown by the
red curve in Fig.\ \ref{Fig:comberr}.  Another important source of
uncertainty is due to uncertainty in the methane absorption
coefficients, which is not easy to characterize.  We assumed that the
uncertainty in $k$ had the form $\epsilon(k) = \alpha k + \epsilon_0$,
where we adopted values of $\alpha = 0.02$ and $\epsilon_0$ = 5$\times
10^{-4}$ km-am.  The 2\% scale-factor component ($\alpha$) is roughly
in agreement with the verification provided by the Descent
Imager/Spectral Radiometer (DISR) measurements within Titan's
atmosphere \citep{Tomasko2008ch4} over the 0.5-0.75 \mum wavelength
range, though larger errors are indicated at longer wavelengths.  The
offset error is even less certain.  KT2009 made changes as large as
0.01 km-am from previous coefficients \citep{Kark1998Icar}, and the
new values, which we use here, are likely to be much smaller, but by
exactly what factor is unclear.

We used a refinement of the approach of \cite{Sro2010iso} in approximating
the effect of methane absorption uncertainty on the spectrum.  We
used a similar crude approximation, that the reflected intensity at any wavelength
could be expressed as a maximum times an exponential in optical depth, i.e.
\begin{eqnarray}
I(\lambda,\mu)&=&I_0(\mu)\exp (-k u/\mu)
\end{eqnarray}
where the optical depth is $k u$, $k$ is the absorption coefficient,
and $u$ is the absorber path amount.  This directly leads to a
fractional error in I/F given by
\begin{eqnarray}
\delta I/I = 1- [I(\lambda,\mu)/I_0(\mu)]^{\alpha+\epsilon_0/k}
\end{eqnarray}
which differs from the model of \cite{Sro2010iso} by using
the maximum $I_0(\mu)$ instead of unity, and by including the
$\epsilon_0/k$ term in the exponent.  That term becomes dominant for
small $k$, and expresses the fact that small absorptions that cause
large I/F depressions lead to large uncertainties because the offset
uncertainty begins to be comparable to the total absorption
coefficient.  However, in this crude form the term becomes unstable
when the absorption is very small and $I/I_0$ is near 1, but where
(because of our crude model) it is not close enough to 1 to make the
$\epsilon_0/k$ term unimportant.  To counteract this instability we
replace $\epsilon_0/k$ with $\epsilon_0/(k+k_\mathsf{min})$, where we
chose $k_\mathsf{min}$=0.01 to avoid excessively large errors for
$\lambda <$ 0.5 \mumx.  The final form of this model is shown as the
\chf uncertainty in Fig.\ \ref{Fig:comberr}. This was treated as normally
distributed with the standard deviation as shown, which is another
crude assumption.  However, it is likely that the overall dominance of
this error source at wavelengths exceeding about 0.6 \mum is
correct.

 Our final combined error estimate was the square root of the sum of
the squares of the three sources.  Although these error sources are
not accurately known or well characterized by our noise model, they
can be roughly validated by comparing, for the best fit over all
models and profiles, the minimum $\chi^2$ value with the expected
value for a perfect fit. Our very best fit over this spectral range
(at this latitude) yielded approximately 339 for $\chi^2$ instead of
the value of 269 predicted by the error model.  This would be the
result of under-estimating the combined errors by 12\%, or by a defect
in the physical model we are using to fit to the observations (at
other latitudes we obtained \chisq values as low as 269).

\subsection{Fitting procedures}
To avoid the need for fitting an imaginary index in the methane layer
(the UTC layer) we fit only the wavelength range from 0.55 \mum to 1.0
\mumx.  To provide a reasonable computational speed while still
sampling a wide range of penetration depths for each spectral band, we
chose a wavenumber step of 118.86 \icm for sampling the observed and
calculated spectrum. This yielded 69 spectral samples, each at four
different zenith angle cosines (0.3, 0.4, 0.6, and 0.8), for a total
of 276 points of comparison. Our compact layer model has seven
adjustable parameters (pressures and optical depths of the UTC and MTC
layers, the optical depth per bar of the stratospheric haze, the
optical depth per bar of the upper tropospheric haze (KT2009 referred
to this as the upper tropospheric layer), and the optical depth of the
LTC layer, leaving 269 degrees of freedom. We fixed the LTC base
pressure to 5 bars and the UTC mean particle radius at 1.2 \mum
because these values were consistently obtained for a variety of fits,
and reducing the number of adjustable parameters improved fit
algorithm performance. Furthermore, the LTC pressure is not well
constrained by the observations because its change can be compensated
by changing the LTC optical depth.  To fit the KT2009 model over the
same range, we followed their approach by adjusting only the four
$d\tau/dP$ (optical depth per bar) values.  We use a modified
Levenberg-Marquardt non-linear fitting algorithm \citep{Sro2010iso} to
adjust the fitted parameters to minimize $\chi^2$ and to estimate
uncertainties in the fitted parameters.  The uncertainty in $\chi^2$
is expected to be $\sim$25, and thus fit differences within this range
are not of significantly different quality.

\section{Application of Spectral Constraints to the Occultation Solutions}

\subsection{Fit results for 5\deg S}
\begin{figure}\centering
\includegraphics[width=3.2in]{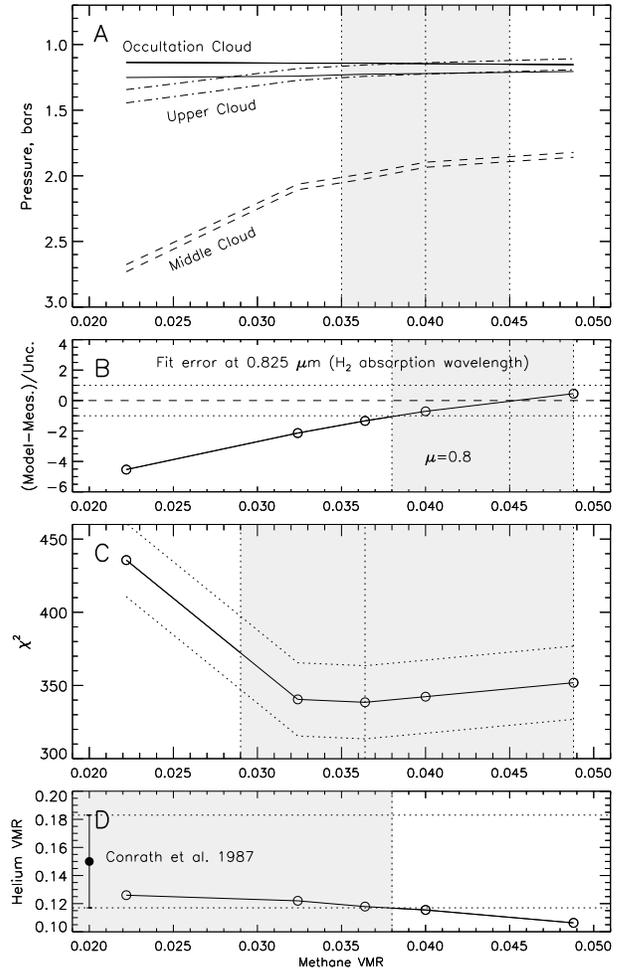}
\caption{Fit results for compact cloud layer models as a function of
the deep volume mixing ratio of methane (points are plotted for models
D1, E1, EF, F1, and G from left to right). A: pressures of the upper
(presumably \chfx) and middle tropospheric cloud models compared to
the cloud layer inferred from our occultation analysis (solid
lines). B: Fit error at 0.825 \mum where hydrogen CIA is prominent.
C: Minimum $\chi^2$ values (solid line) for fits to spectra at 4
zenith angles and 69 wavelengths from 0.55 to 1.0 \mumx, with dotted
lines indicating the 1$\sigma$ uncertainty range expected for \chisqx.
D: He volume mixing ratio (line) compared to the \cite{Conrath1987JGR}
value (symbol with error bars). The grayed areas indicate rough
regions of uncertainty/acceptability.}
\label{Fig:cloudpvch4}
\end{figure}

The fit results for the compact model for profiles D1, E1, EF, F1, and
G are given in Table \ref{Tbl:fitsum} and key results are plotted as a function of
methane volume mixing ratio in Fig.\ \ref{Fig:cloudpvch4}.
The parameters of the aerosol model are very well constrained by the
observations, with pressures constrained to a fraction of a percent
and optical depths usually to within 5\%. 
In Fig. \ref{Fig:cloudpvch4} we show several results that can be used
to constrain the methane mixing ratio at 5\deg S: (A) the pressure of
the upper tropospheric cloud (UTC) in comparison with the occultation
cloud pressure, (B) the overall quality of the spectral fit, (C)
the fit error at 0.825 \mumx, and (D), the He/H$_2$ ratio.

The results for cloud pressure (Fig. \ref{Fig:cloudpvch4} A) show that
the upper compact cloud (modeled as methane) is in best agreement with
the occultation pressure range at a methane mixing ratio of 4.0\%, but
the match is still fairly close down to a mixing ratio near 3.5\%.  An
even stronger constraint on the \chf mixing ratio is the fitting error
in the region near 0.825 \mum (Fig. \ref{Fig:cloudpvch4}B), where
hydrogen CIA exceeds methane absorption. If there is too much methane
assumed in our model, then model cloud particles will need to move
upward to compensate. That will place them relatively further above
the absorption of hydrogen, and where the effects of hydrogen can be
seen (as at 0.825 \mumx) the model will appear too bright relative to
the observations.  Where the assumed \chf mixing ratio is too low (as
for Model D1 at 5\deg S) the model I/F will appear be too low at
0.825 \mumx.  This constraint clearly favors a mixing ratio of 4.5\%
with an uncertainty of roughly 0.7\%.  The overlap of the first two
uncertainty ranges is 3.8-4.5\%.  The third constraint is overall fit
quality (Fig. \ref{Fig:cloudpvch4} C). The best spectral fit (judged
from the minimum $\chi^2$ in the middle panel) is for a methane mixing
ratio near 3.6\%, but the fit is nearly as good over a wide range from
3-4.9\%.  This is a relatively weak constraint that is easily
compatible with the previous stronger constraints, which favor a
mixing ratio of 4\%.  This makes our Model F1 profile the preferred
profile, even though it slightly exceeds the \cite{Conrath1987JGR}
uncertainty limits of the He VMR, as shown in Fig.\
\ref{Fig:cloudpvch4}D.  We consider E1, EF, and F1 profiles plausible
candidates for further analysis.  Our preferred model (F1) leads to a
compact methane condensation cloud very close to pressure level
expected from occultation results and it is also a model that provides
an excellent fit to the STIS spectral observations, especially in the
key region where hydrogen CIA is significant.

\begin{table*}\centering
\caption{Best-fit parameters for compact cloud layer models at 5\deg S.} 
\vspace{0.15in}
\begin{tabular}{r c c c c c}
OCCULTATION PROFILE: &  \hspace{0.15in}   D1 \hspace{0.15in}    &  \hspace{0.15in}   E1\hspace{0.15in}    &   \hspace{0.15in}  EF \hspace{0.15in}    &  \hspace{0.15in}  F1 \hspace{0.15in}    &  \hspace{0.15in}  G\hspace{0.15in}  \\[0.05in]
\hline
\\[-.1in]
KEY PROFILE PARAMETERS:\\[0.05in]
Cloud Top, bar      &       1.136 &  1.142 & 1.142&  1.146&  1.153 \\[0.05in]
Cloud Bottom, bar      &       1.251 &  1.241 & 1.226&  1.221&  1.205 \\[0.05in]
Deep Methane VMR&      2.22\%&  3.24\%& 3.64\%& 4.00\%& 4.88\% \\[0.05in]
AEROSOL PARAMETERS:\\[0.05in]
$(d\tau/dP)_\mathsf{Strat.H}$, bar$^{-1}$ & 0.329&  0.208& 0.178&  0.158&  0.136 \\[0.05in]
$(d\tau/dP)_\mathsf{UTH}$, bar$^{-1}$ &  0.01&  0.032& 0.035&  0.038&  0.031\\[0.05in]
$P_{\mathsf{UTC}}$, bar&  1.45&  1.27& 1.24&  1.23&  1.19\\[0.05in]
$P_{\mathsf{MTC}}$, bar&  2.54&  1.80& 1.72&  1.68&  1.60\\[0.05in]
$P_{\mathsf{LTC}}$, bar&    5.&       5.&      5.&       5.&       5.\\[0.05in]
{\Large$\tau$}$_\mathsf{UTC}$  &   0.444&  0.328& 0.324&  0.322&  0.324 \\[0.05in]
{\Large$\tau$}$_\mathsf{MTC}$  &   1.33&  1.16& 1.23&  1.28&  1.42 \\[0.05in]
{\Large$\tau$}$_\mathsf{LTC}$  &     0.01&  2.30& 2.92&  3.45&  4.75\\[0.05in]
{\Large$r$}$_\mathsf{Strat.H}$, \mum &   0.1& 0.1& 0.1& 0.1&  0.1\\[0.05in]
{\Large$r$}$_\mathsf{UTC}$, \mum &   1.2&  1.2& 1.2&  1.2&  1.2\\[0.05in]
FIT QUALITY:\\[0.05in]
{\large$\chi^2$(total)}        & 435.6&   340.5&  338.5&   342.3&   351.9 \\[0.05in]
{\large$\chi^2$(0.825 \mumx)}        & 47.7&   11.0&  5.1&   2.2&   0.6 \\[0.05in]
Fit error at 0.825 \mumx, $\mu$=0.8  & -4.54&   -2.14&  -1.34&   -0.71&   +0.54 \\[0.05in]
\hline\\[-0.1in]
\end{tabular}\label{Tbl:fitsum}
\parbox{5in}{Note: The uncertainty in \chisq is $\sim$25 and thus
fits differing by less than this are not of significantly different
quality. The stratospheric haze base extends upward from 100 mb, the
upper tropospheric haze (UTH) extends from 900 mb to 100 mb, The upper
tropospheric cloud (UTC) and middle tropospheric cloud (MTC) extend
from the bottom pressures, tabulated here, to 0.93 times those
pressures.  The lower tropospheric cloud is bounded by the tabulated
base pressure and 0.98 times that pressure. For these fits, the
particle radii of the Mie layers were held fixed, as was the pressure
of the lower tropospheric cloud. The fit error at 0.825 \mum is the
ratio of (model-measured) to estimated uncertainty.}
\end{table*}

\begin{figure}\centering
\includegraphics[width=3.2in]{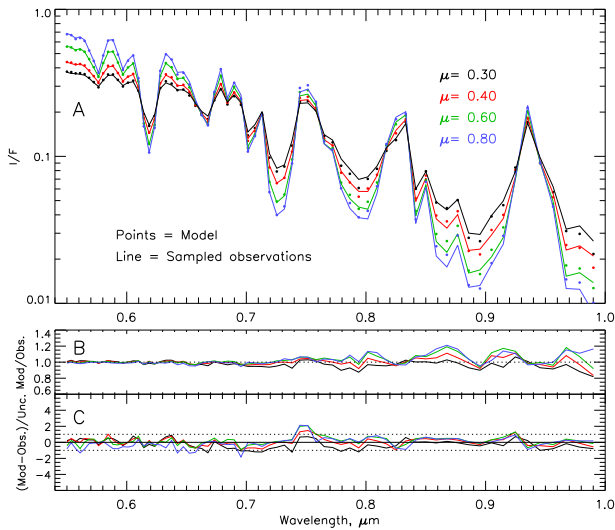}
\caption{A: STIS 5\deg S spectra (solid lines) at four zenith angles, sampled at a wavenumber
spacing of 118.86 \icmx, compared to model calculations (points) for the model that
best fits the spectrum when the F1 profile of temperature and methane mixing
ratio is assumed. B: Ratio of Model to observed I/F values. C: Difference of model
and observed values divided by the expected uncertainty at each wavelength sample. Note that
the larger fractional errors that occur for $\lambda > $0.78 \mum are within the
expected range of uncertainties.}
\label{Fig:F1fit}
\end{figure}

The best compact cloud model fit to STIS spectra at 5\deg S using the
F1 profile (shown in Fig.\ \ref{Fig:F1fit}) yielded an uncorrected
$\chi^2$ of 342, which is significantly better than the best $\chi^2$
values of 409, 426, and 458 obtained by our fit of the KT2009 model to
E1, EF, and F1 profiles respectively.  Thus, we don't have to give up
anything in fit quality to obtain the compatibility between aerosol
models and occultation models. In fact, the fit quality improves, as
one would expect with two more adjustable parameters available.

Our presumed methane cloud has an effective particle radius near 1.2
\mum and a rather small optical depth, only about 0.32 at $\lambda=$
0.5 \mumx, which about half the value reported by \cite{Rages1991} for
their low latitude cloud, but close to the 0.4 value of
\cite{Baines1995} for a disk average that weighted high latitudes
most.  The middle tropospheric cloud is considerably thicker, with an
optical depth of 1.23 for the optimum methane profile.  The much
deeper lower tropospheric cloud is generally even thicker, except for
the model D1 profile for which this cloud is not even needed.

\subsection{Layer contributions}

\begin{figure*}\centering
\hspace{-0.2in}
\includegraphics[width=3.2in]{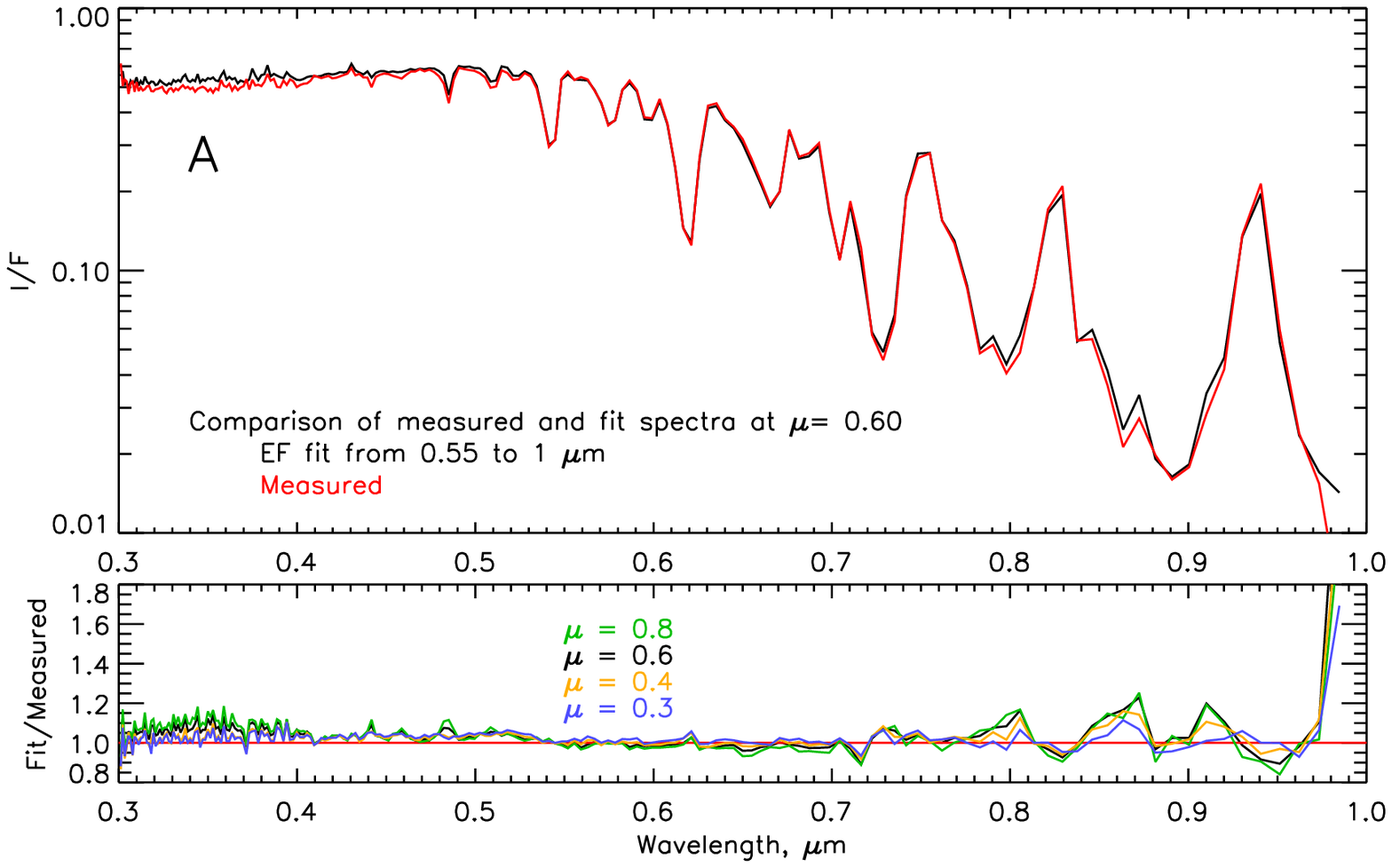}
\includegraphics[width=3.2in]{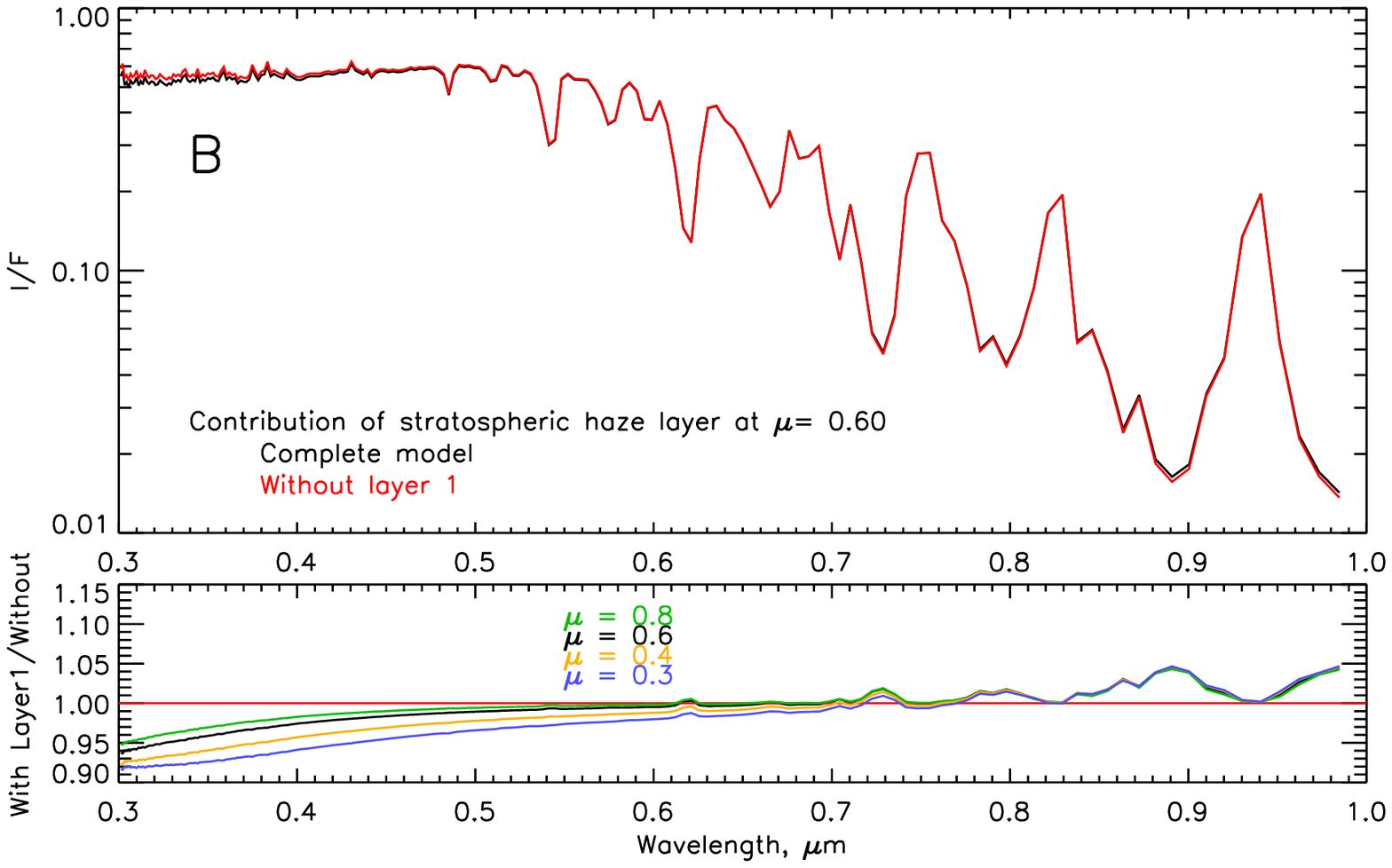}\\
\hspace{-0.2in}
\includegraphics[width=3.2in]{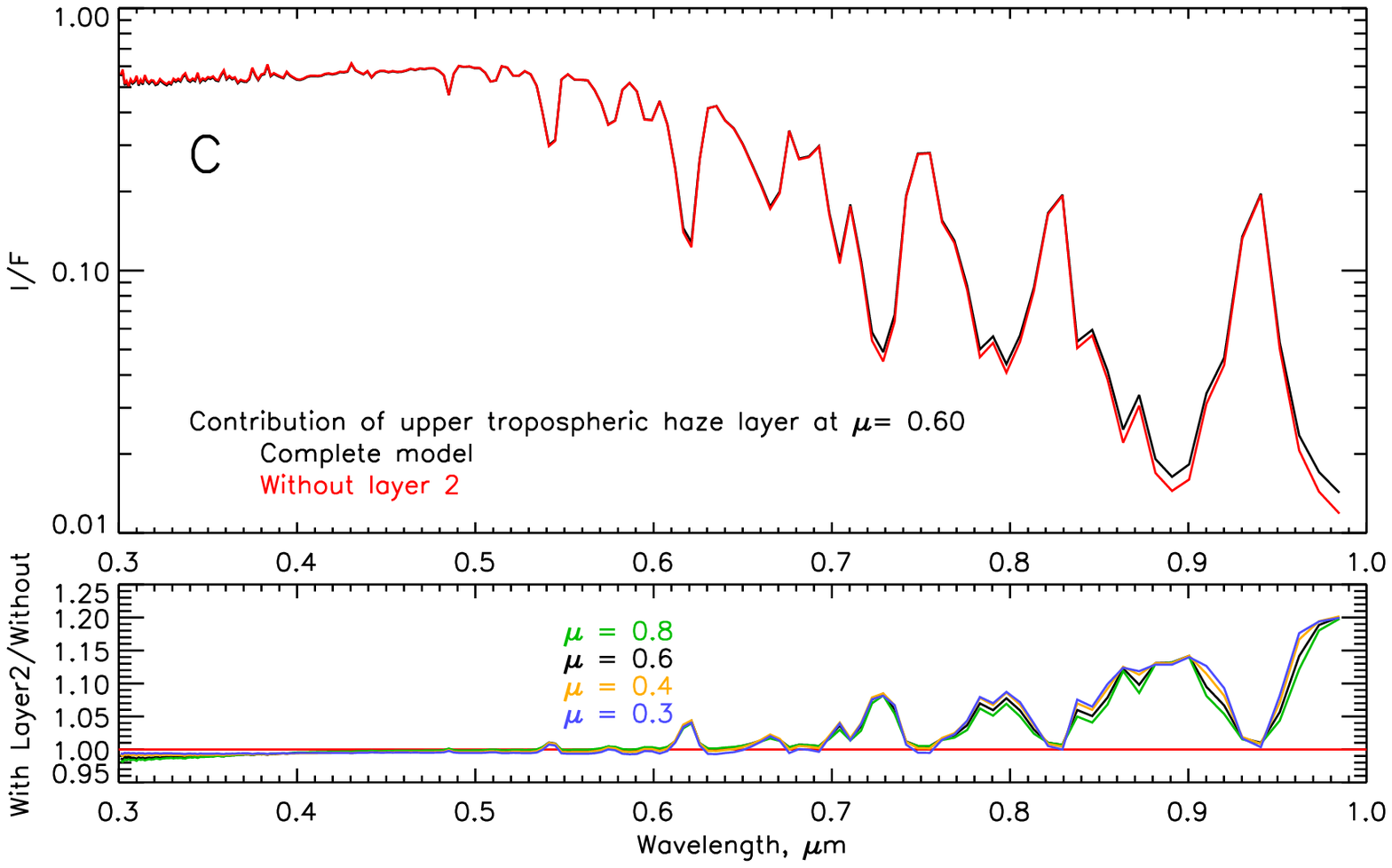}
\includegraphics[width=3.2in]{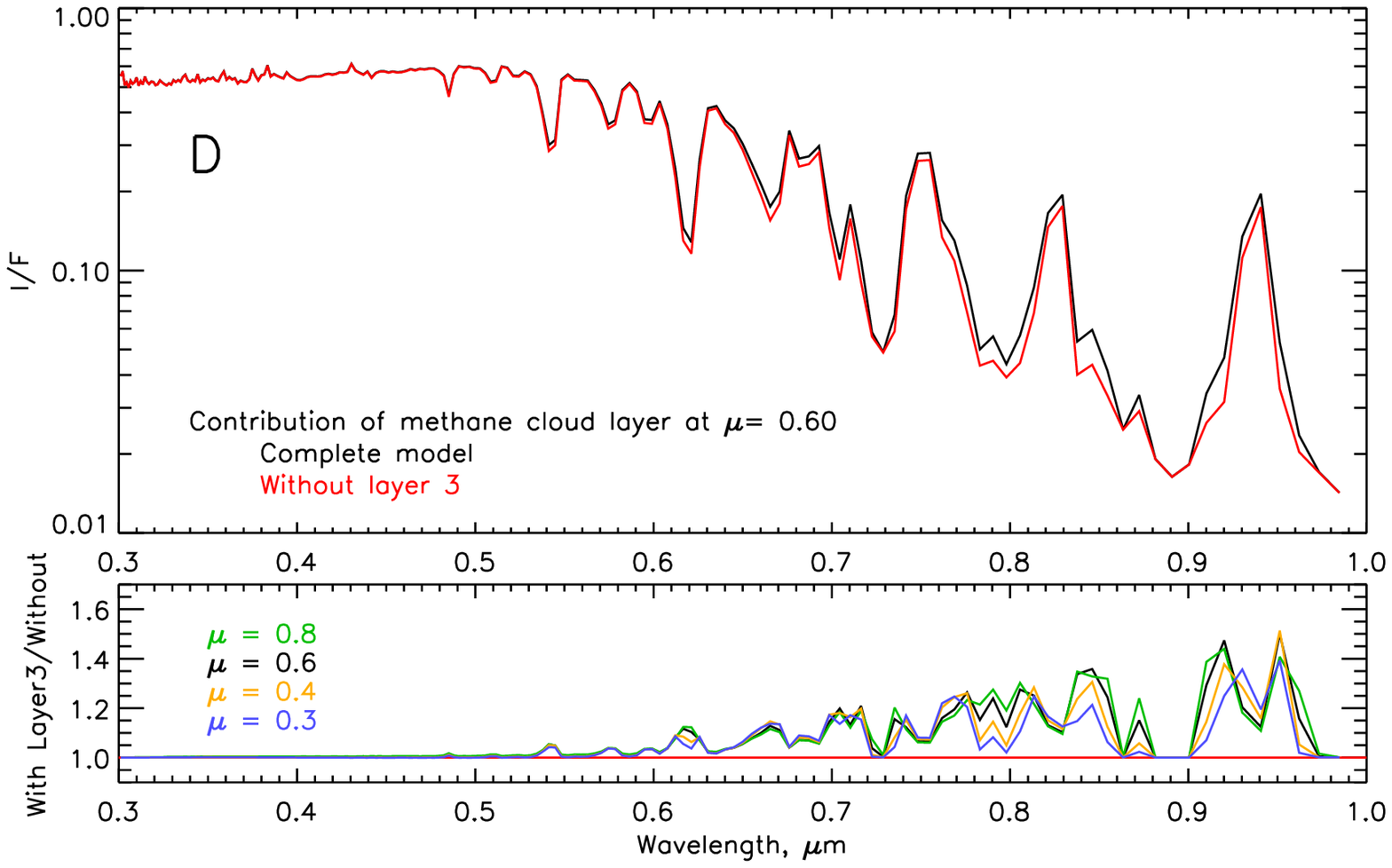}\\
\hspace{-0.2in}
\includegraphics[width=3.2in]{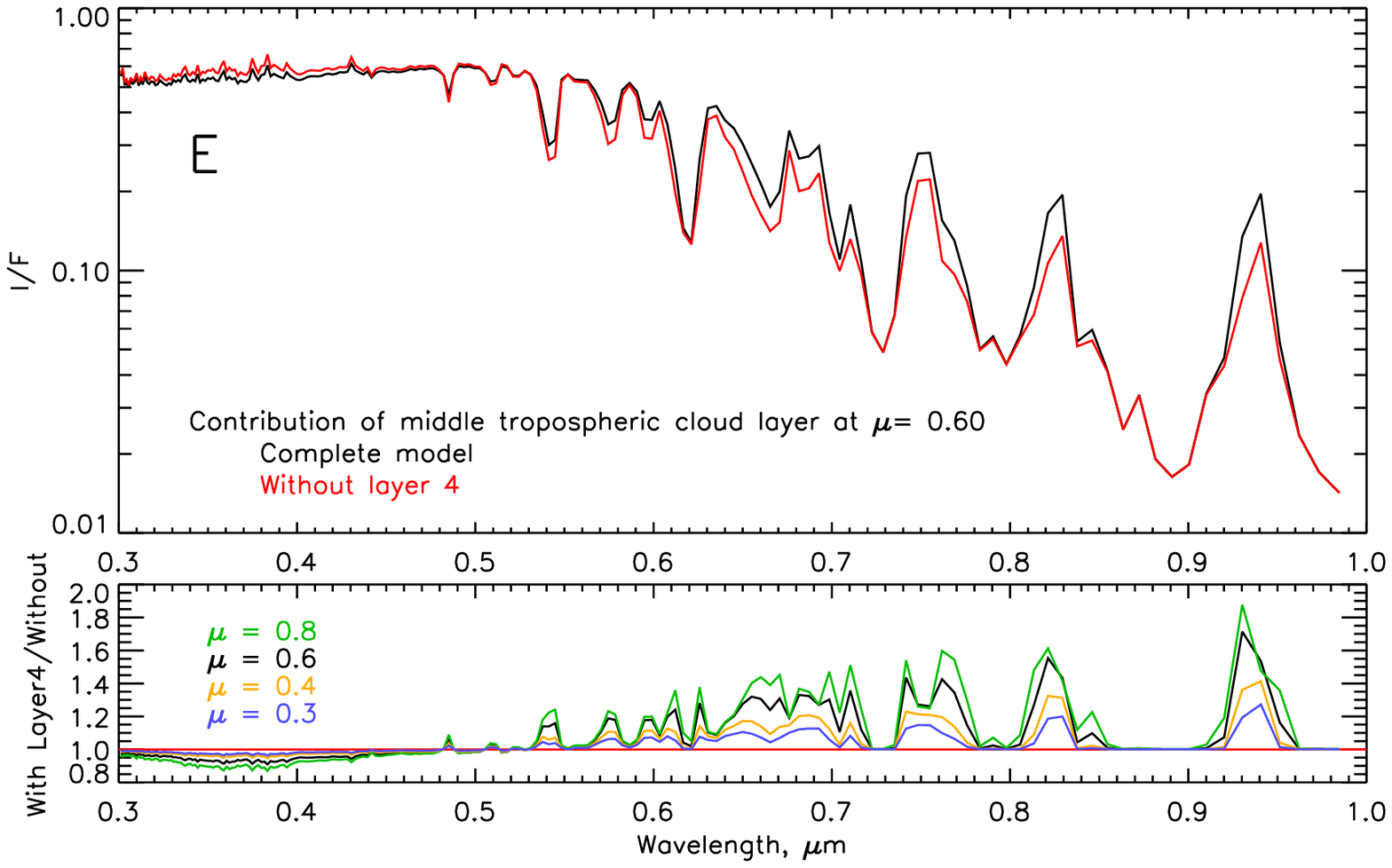}
\includegraphics[width=3.2in]{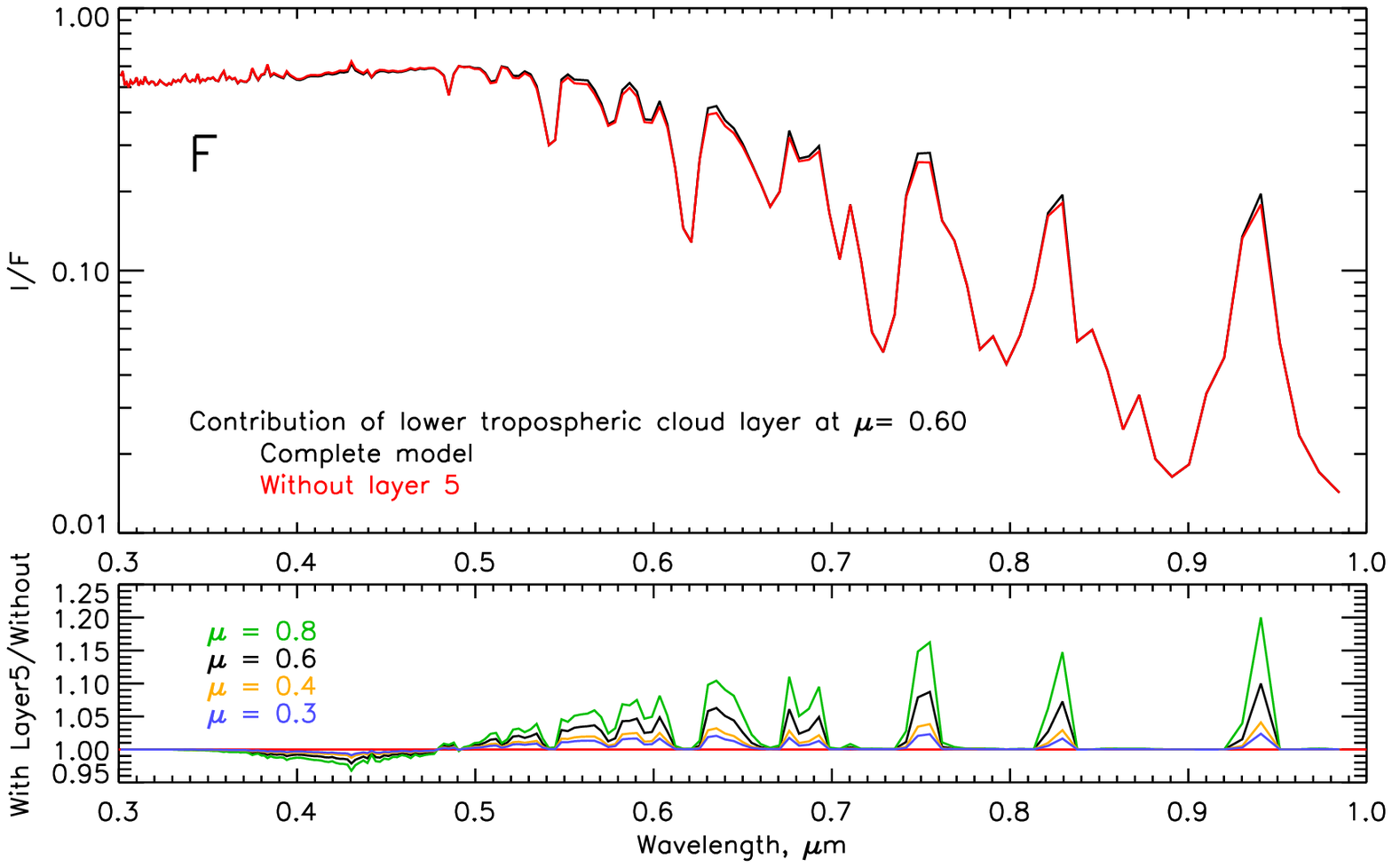}\\
\caption{A: Comparison of measured STIS spectrum (red) at 5\deg S with
a compact model fit (black), with spectra shown at a solar zenith
angle cosine of 0.6 and ratios shown for all four cosine values); B-F:
Comparisons of the model spectrum with spectra for same model with
layers 1-5 removed respectively.  Layer 3 (D) is the methane cloud
layer. Although these spectral sensitivities were calculated for the EF
profile, they also closely represent results for E1 and F1 profiles.}
\label{Fig:contributions}
\end{figure*}

 The relative roles played by our five model layers in creating the
observed spectral characteristics are illustrated in Fig.\
\ref{Fig:contributions}B-F, which displays the effect of removing each
layer from the model spectrum. The distinctly different contributions
of the various layers is what allows the model parameters to be so
well constrained by the observations.  We see in B that the
stratospheric haze (layer 1) serves to reduce the I/F for $\lambda <
0.6$ \mum and provide a slight (5-10\%) boost to the I/F in the center
of the deep absorption bands at longer wavelengths.  As shown in C,
the upper tropospheric haze layer (layer 2) makes a similar
contribution but without as much shortwave absorption.  The methane
cloud (layer 3, panel D) contributes almost nothing at wavelengths
less than 0.55 \mum but provides a significant contribution to minima
in the intermediate absorption bands and especially to the shoulders
of the strong absorption bands. The effect of the middle tropospheric
cloud (layer 4, panel E) is seen mainly in the strong contributions to
the I/F peaks, and also in contributing a small absorption peaking
near 0.4 \mumx.  The lower tropospheric cloud (layer 5, panel F)
contributes mainly to the longer wave window regions, and is unique in
providing key contributions at 0.55 and 0.59 \mumx.  Note that,
although the model fit was only constrained by measurements from 0.55
to 1.0 \mumx, it provides a relatively good fit to the shorter
wavelengths as well (panel A). The discrepancies between model fit and
measurements between 0.3 and 0.4 \mum are only about 10\% and could be
largely removed by slight increases in the lower tropospheric cloud
absorption and slight decreases in the stratospheric haze absorption
in this region. It might also be possible to introduce some of the
extra absorption needed at small zenith angles into the methane layer
itself, as suggested by \cite{Rages1991}.

\subsection{Comparison of vertical structures}

\begin{figure*}\centering
\includegraphics[width=5in]{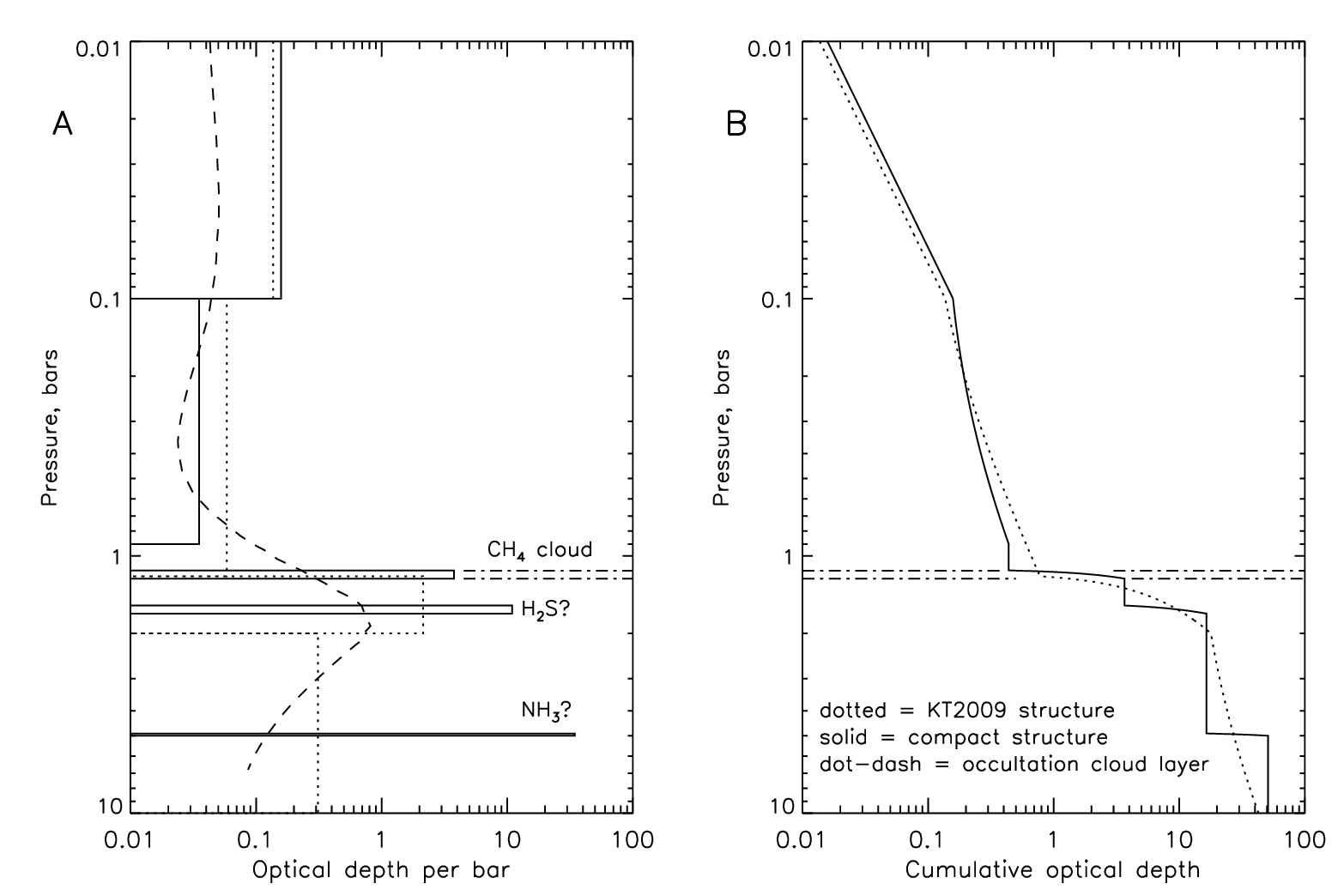}
\caption{Optical depth per bar (A) and cumulative optical depth (B)
for a vertically diffuse KT2009 model (dotted) and out compact layer
model (solid) fit to 5\deg S STIS spectra using the F1 vertical
profile of temperature and methane mixing ratio. For the KT2009 model we
fit the four optical depths per bar, and used the KT2009 values for
pressures and particle scattering properties. For our model we fit two
optical depths per bar for the upper two layers and the pressures and
optical depths of the three compact layers, except for the fixed 
5 bar pressure we used for the bottom layer. The dashed line in A is
optical depth per bar at 1.6 \mum from Fig. 5 of \cite{Irwin2010Icar} with
pressures scaled by the factor 1.8/2.6 to account for the decrease
indicated in their Fig.\ 8, when they used a vertical methane profile
similar to the L87 Model F.}
\label{Fig:cumod}
\end{figure*}

The vertical structure of our compact cloud model and the diffuse
vertical structure of the KT2009 model are compared directly in Fig.\
\ref{Fig:cumod}, which displays optical depth per unit pressure versus
pressure (A) and cumulative optical depth versus pressure (B). There
is a very large difference in the vertical distribution of cloud
particles, with our compact model providing high opacity
concentrations in narrow pressure ranges. But when compared on the
basis of cumulative opacity we see that the KT2009 model looks like a
vertically averaged version of our results.  The compact layer with
the strongest effect on the CCD spectrum is our layer 4 (MTC), as
shown in Fig.\ \ref{Fig:contributions}. The dashed line in Fig.\
\ref{Fig:cumod}A is for optical depth per bar at 1.6 \mum from Fig. 5
of \cite{Irwin2010Icar} with pressures scaled by the factor 1.8/2.6 to
account for the decrease indicated in their Fig.\ 8, when they used a
vertical methane profile similar to the L87 Model F. Their relative
vertical distribution of opacity is crudely consistent with that of
the other models, when vertical smoothing is taken into account.
However, the quantitative values of their opacities per unit pressure
are hard to compare with our (or KT2009) results because of very
different assumptions made about particle scattering properties,
including the extreme assumption that particle scattering properties
were independent of latitude and altitude.

While the top compact layer in our model is consistent with
condensation of methane, the composition of the deeper layers is
highly uncertain.  Possible parent gases are H$_2$S and NH$_3$, but
condensation layers at 1.7-3 bars would imply strong depletions from
their expected deep mixing ratios, which is plausibly an effect of the
formation of a much deeper cloud of NH$_4$SH particles, as discussed
by \cite{DePater1989Icar82} and \cite{Fegley1991}.

\section{Latitude dependence of methane on Uranus}

We first show that the occultation solution that is most consistent
with low latitude spectra is not consistent with high-latitude spectra and
then discuss how that profile can be modified in physically reasonable
ways and which modifications provide the best fits at high latitudes.

\subsection{Fit results assuming latitude-independent methane}

\begin{figure}\centering
\includegraphics[width=3.2in]{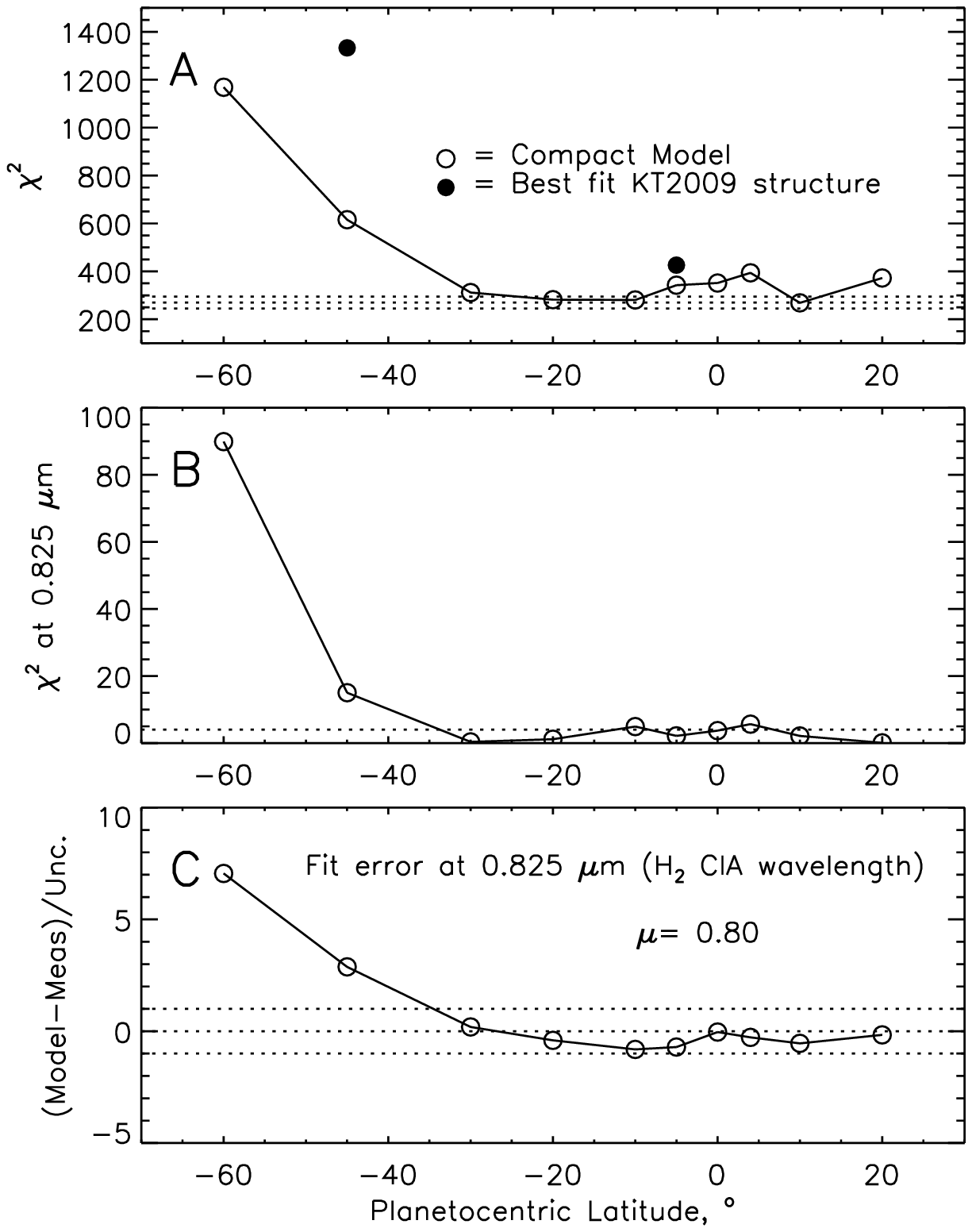}
\caption{Spectral fit quality measured by \chisq over the entire range
of the fit (A) and \chisq for the 4 view angles at the 0.825 \mum H$_2$
CIA wavelength (B), and fit error at 825 nm and $\mu$=0.8 (C), all as
a function of planetocentric latitude, using the F1 temperature and
methane profiles. Filled circles in A denote results of fitting the
KT2009 structure model. The horizontal dotted lines in A and C indicate
expected (central) and 1$\sigma$ uncertainty bounds.}
\label{Fig:laterr}
\end{figure}

EF and F1 model temperature and methane profiles yielded high quality
spectral fits not only at the occultation latitude, but also for
latitudes from 30\deg S to 20\deg N, as indicated in Fig.\
\ref{Fig:laterr}A for the F1 fits. Over this latitude range the fit
quality was generally very high, reaching \chisq values close to those
expected from the error model, and about half were notably better than
we obtained at 5\deg S.  However, between 30\deg S and 45\deg S fit
quality began to deteriorate dramatically, with \chisq growing to
$\sim$1100 by 60\deg S, which is four times the expected value.  At
the same time, at 0.825 \mumx, where hydrogen CIA is an important
contributor, the \chisq contribution (Fig.\ \ref{Fig:laterr}B) and the
signed fit error (Fig.\ \ref{Fig:laterr}C) remained relatively
flat over the 30\deg S to 20\deg N region, but grew dramatically from
30\deg S to 60\deg S, with the signed error reaching 7 times the
expected uncertainty and the \chisq contribution reaching more than
40 times the expected value.  Thus, there is little question that the
methane mixing ratio at high latitudes is quite different from what it
is at middle latitudes, a result already established by KT2009.  To
get a more realistic picture of the high latitudes, we need to
estimate what the actual methane mixing ratio profile is at these
latitudes. In the next section we discuss plausible ways to modify our
baseline (F1) profile so that we can investigate this issue.

\subsection{Construction of depleted methane profiles}

The occultation places no direct constraints on the temperature or
methane profiles at other latitudes.  However, there are some
reasonable physical constraints to consider. First, the lack of
significant latitudinal gradients in the temperatures inferred from
Voyager 2 IRIS observations for P$> \sim$150 mb \citep{Hanel1986Sci}
suggests that the occultation thermal profile is a good approximation
at all latitudes on Uranus. A second constraint has to do with an
extension of the relationship between the vertical wind shear and
horizontal temperature gradients.  The variation of mixing ratio with
latitude, assuming the same pressure-temperature structure at all
latitudes (as assumed by KT2009 as well), leads to a variation in density on
constant pressure surfaces, which is similar in effect to horizontal
temperature gradients.  Both kinds of gradients lead to vertical wind
shear \citep{Sun1991}, a consequence of geostrophic and hydrostatic
balance.  If such latitudinal variations occurred through great
atmospheric depths, this would lead to great differences in cloud
level winds and conflict with the observed wind structure of
Uranus. Thus we expect that where methane is depleted, it is not
depleted over great depths.  And, as indicated by KT2009, the spectral
observations do not require that methane depletions extend to great
depths. In fact, we will later show that the spectral constraints
favor relatively shallow methane depletions.

\begin{figure}\centering
\includegraphics[width=3in]{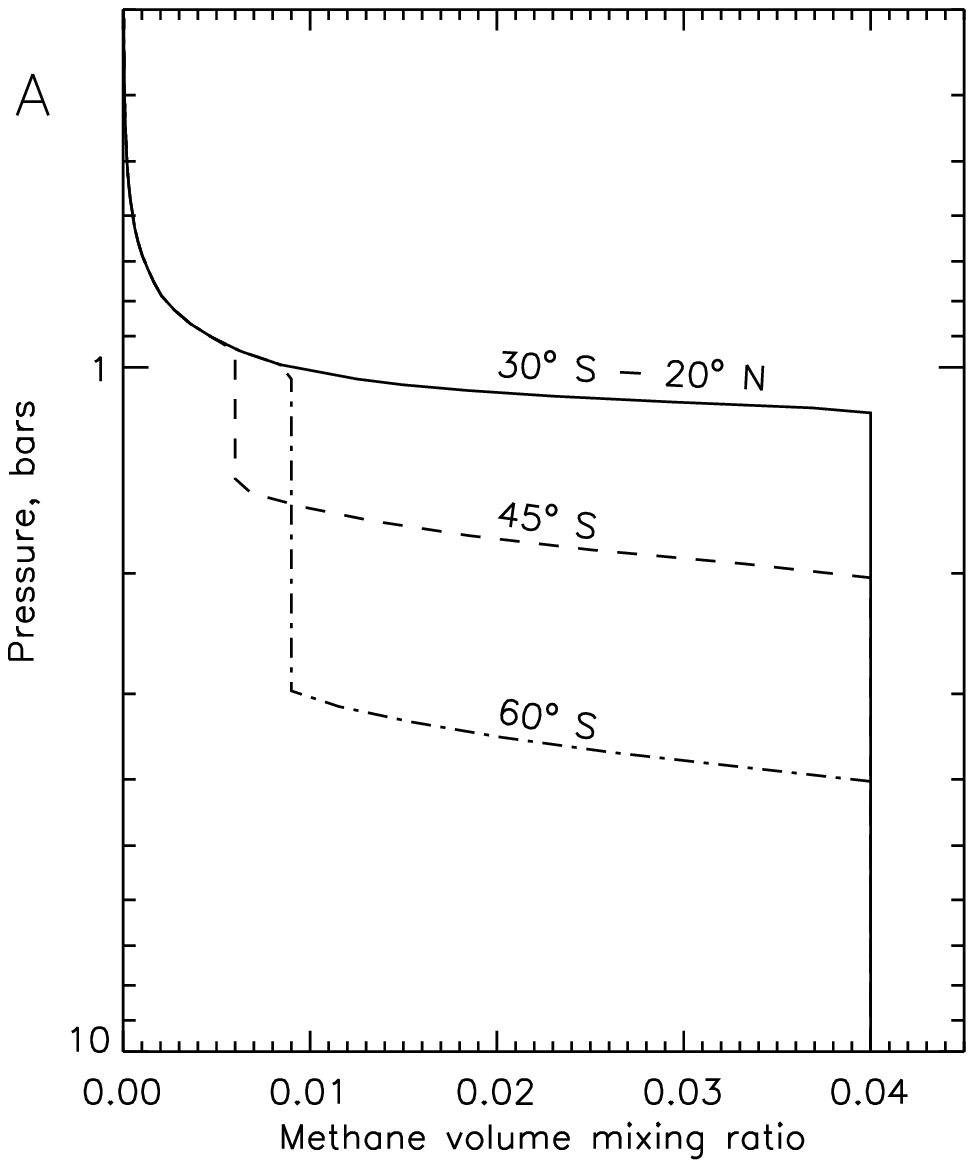}
\includegraphics[width=3in]{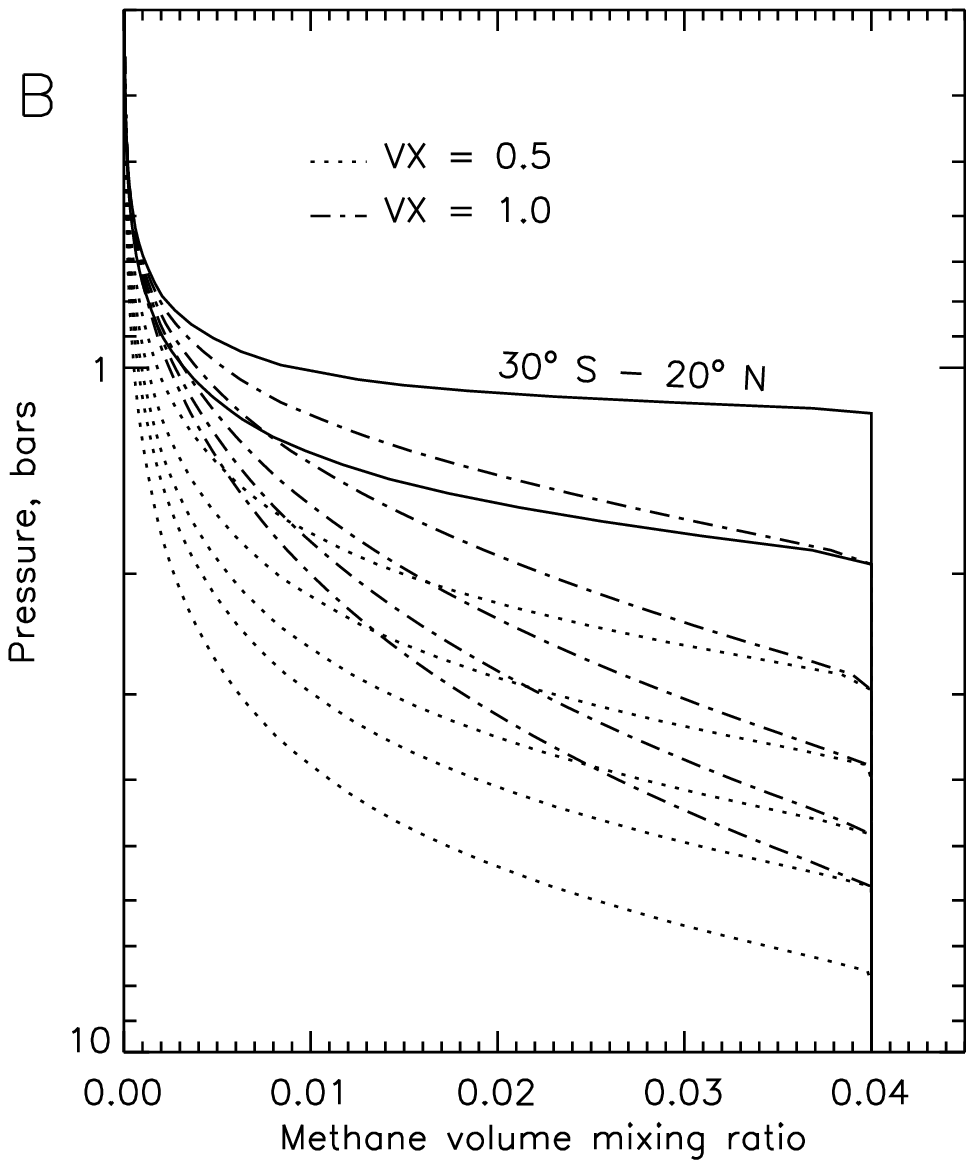}\par
\caption{Sample profiles of depleted methane used to fit high-latitude STIS spectra. The undepleted
base profile labeled by its valid latitude range of 30\deg S to 20\deg N is from Model F1. The
profiles in A provided the best fits for the indicated latitudes. None of the B profiles
yielded any better fits (see text).}
\label{Fig:depletions}
\end{figure}

For trial purposes we created several types of modifications of the F1
profile.  We first defined an upper tropospheric mixing ratio and a
set of transition pressures, $P_1$ and $P_2$.  for $P>P_2$ we kept the
mixing ratio at the F1 deep value of 4\%. For $P<P_1$ we set the
mixing ratio to the upper level value, except where that value
exceeded the F1 value above the methane condensation level, at which
point we reverted to the F1 model profile.  Between the two transition
pressures we interpolated between the upper troposphere and deep
values. Sample depleted profiles of this type are shown in Fig.\
\ref{Fig:depletions}A. These profiles are consistent with methane
condensation at somewhat lower pressures than for the F1 profile.  A
depleted profile of this type might be created by descent of the
low-mixing-ratio gas from slightly above the 1.2 bar level downward to
higher pressures, but to limited depths.

We also considered ``proportionally descended gas'' profiles in which
the model F1 mixing ratio profile $\alpha (P)$ was dropped down to increased pressure levels
$P'(\alpha)$ using the equation
\begin{eqnarray}
 P' = P\times [1 + (\alpha/\alpha_{d})^{vx}(P_{d}/P_{cb}-1)]\\\nonumber
  \mathrm{for}& P_{tr} <P<P_{d},
\end{eqnarray}
where $P_{d}$ is the pressure depth at which the revised mixing ratio $\alpha'=\alpha(P')$
equals the uniform deep mixing ratio $\alpha_{d}$ , $P_{cb}$ is the cloud bottom
pressure (which is where the unperturbed profile departs from the
uniform deep value), $P_{tr}$ is the tropopause pressure (100 mb),
and the exponent $vx$ controls the shape of the profile between
100 mb and $P_{d}$.  A sampling of profiles of this type are shown
in Fig.\ \ref{Fig:depletions}B.  The profiles with $vx=1$ are similar
in form to those adopted by \cite{Kark2011nep}.  A profile of this
type is consistent with the descending gas
beginning at lower pressures than for our initial set of models.
In both cases we would expect methane condensation to be inhibited by
the downward mixing of upper tropospheric gas of low methane mixing
ratios.  How these various types of profiles affected fits to the 45\deg S and
60\deg S spectra is discussed in the next section.

\subsection{Constraining methane profiles at 60\deg S and 45\deg S}

\begin{figure*}\centering
\hspace{-0.2in}
\includegraphics[width=3.2in]{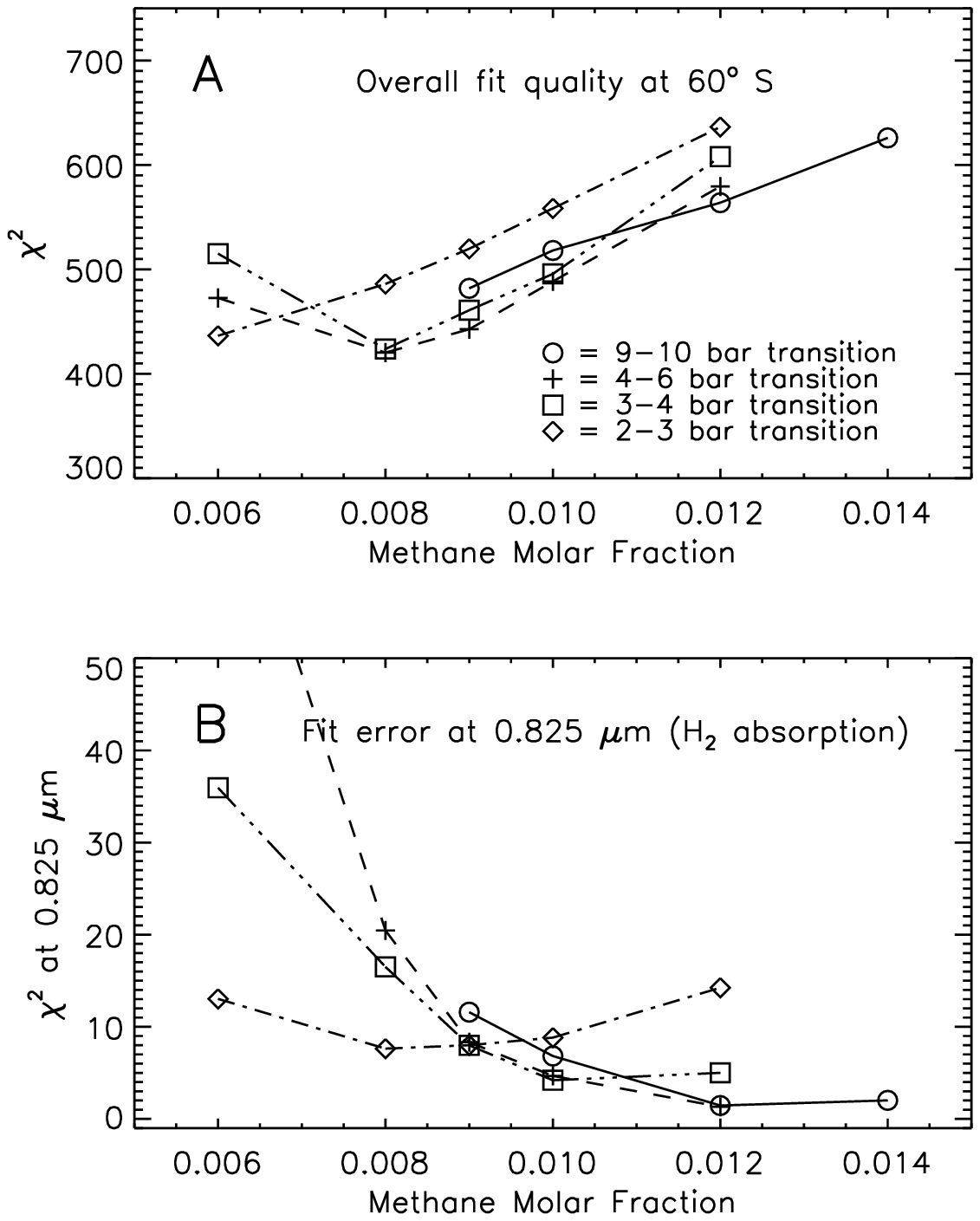}
\hspace{-0.2in}\includegraphics[width=3.2in]{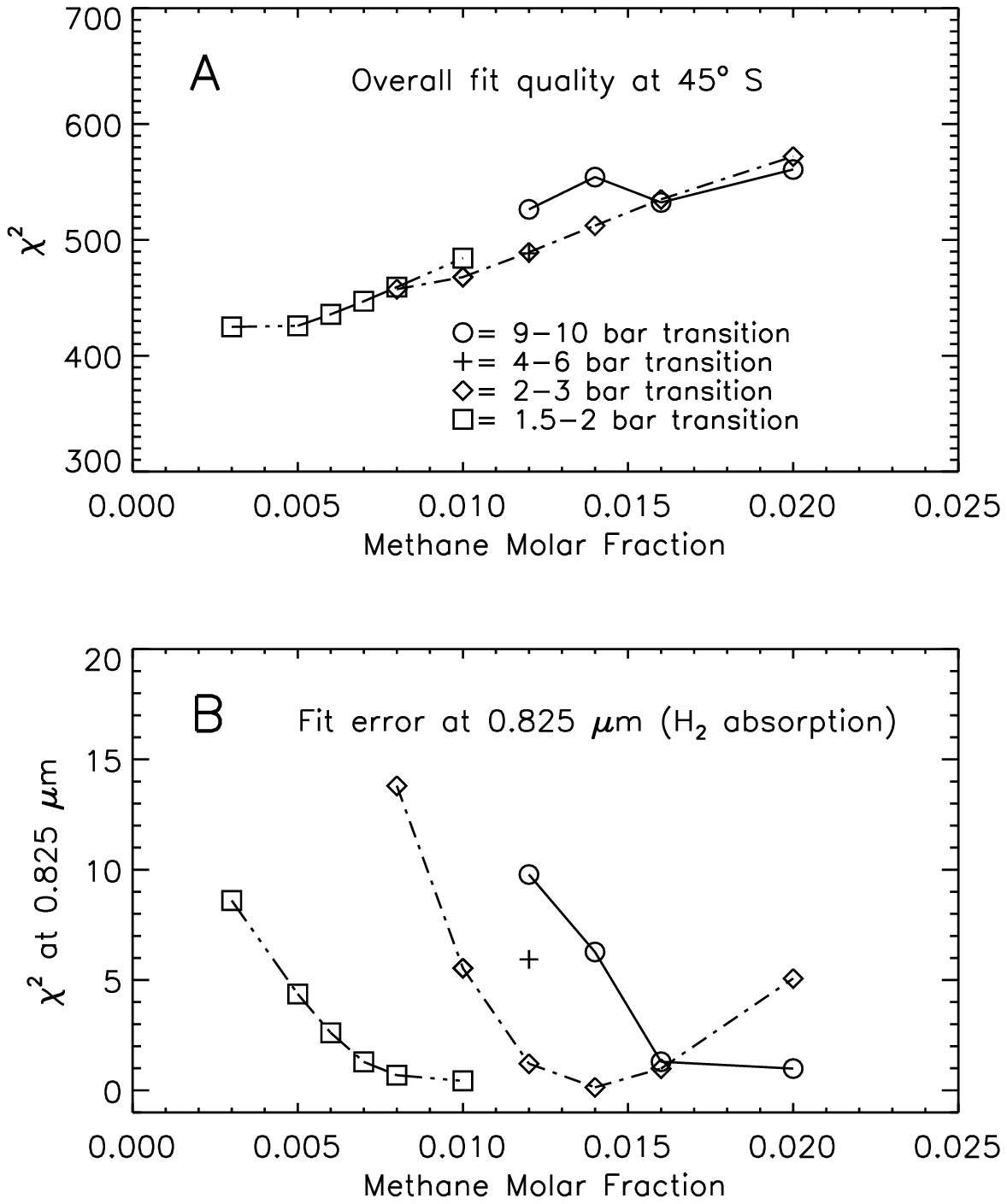}
\caption{Overall fit quality (A) and fit quality at 0.825 \mum (B) versus
depleted mixing ratio for
planetocentric latitudes 60\deg S (Left) and 45\deg S (Right). The transition
pressures between upper depleted regions and the deep region at 4\% are
noted in the legend.}
\label{Fig:45-60fitqual}
\end{figure*}

We used our 5-layer model, allowed adjustment of all 9 parameters, and
tried \chf mixing ratios as low as 0.3\% in the depleted region, and
transition depths ranging from 1.5-2 bars to as deep as 9-10 bars. The
bottom of the transition region is where the F1 deep mixing ratio of
4\% is reached.  The results were evaluated by comparing their overall
fit quality (how low their \chisq values were) and how big their
fitting errors were at the key wavelength of 0.825 \mumx, where
H$_2$ CIA is an important contributor.  The results are summarized
in Fig.\ \ref{Fig:45-60fitqual} for depletion depths of 1.5-2 bars
(only for 45\deg S), 2-3 bars, 4-6 bars, and 9-10 bars.  At neither
latitude is it possible to obtain a good fit with deep depletions that
extend to 9-10 bars.  Although a low hydrogen fitting error can be
obtained from deep depletions, the overall fit quality is much better
for shallow and very strong depletions.  At both latitudes lower
mixing ratios lead to better fits, as long as the depletion depth is
reduced as well. The minimum hydrogen error moves to lower methane
mixing ratios as the depletion depth is decreased and the depletion
factor is increased. It is apparent that the best combination of
overall and CIA fits requires very low methane mixing ratios, of the
order of 1\% or even less, and shallow depletion depths, down to only
1.5-2 bars for 45\deg S, and perhaps 2-3 bars at 60\deg S, which are
illustrated in Fig.\ \ref{Fig:depletions}A.  These perturbations of
the base profile are of such a small depth that the vertical wind
shear caused by the resulting horizontal density gradients would not
likely produce significant perturbations of the zonal wind field,
though exact calculations would be needed to verify that.  The
depletion of methane at high latitude helps to explain the difference
between prior estimates of 4\% by \cite{Rages1991} and 1.1-2.3\% by
\cite{Baines1995}, the latter weighting high latitudes more than the former.

A limited fitting exploration using more physically appealing methane
profiles of the type shown in Fig.\ \ref{Fig:depletions}B did not
yield significantly better fits than those shown in Fig.\
\ref{Fig:depletions}A. We did find constraints on the parameters of
such profiles, however, with $vx=1$ yielding better fits than either
$vx=0.5$ or $vx=2$, and $P_{pd}=2-3$ bars preferred at 45\deg S and
$P_{pd}=4-6$ bars preferred at 60\deg S.

\begin{figure}\centering
\hspace{-0.1in}\includegraphics[width=3.5in]{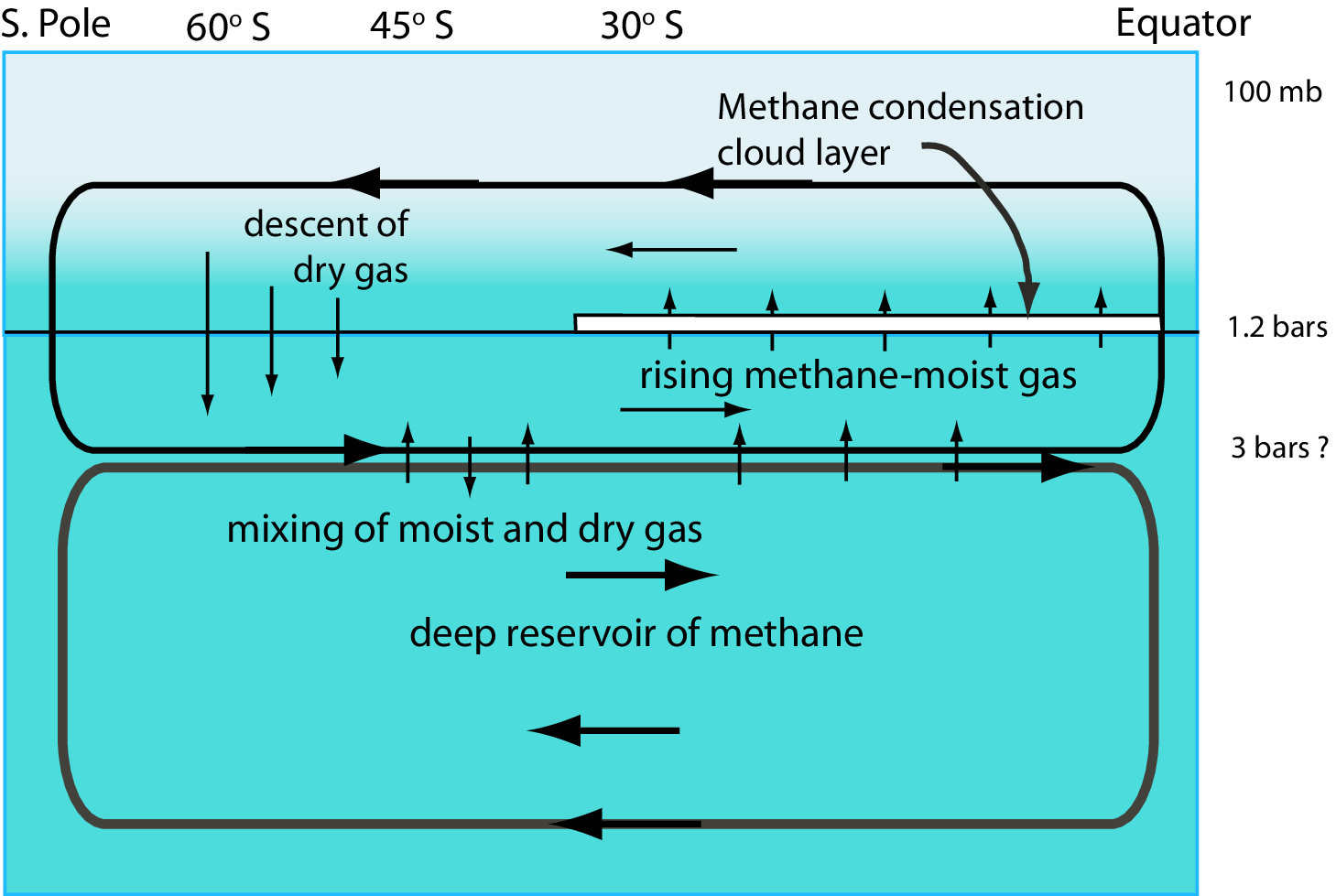}
\caption{Schematic of methane depletion through condensation at
low latitudes as a drying process, poleward transport, then descent
of dry gas, finally mixing with moist gas on the return flow. How
far beyond 60\deg S the depletion extends towards the pole, or whether northern
high latitudes are depleted is unknown. Also unknown is how deep the return
flow extends and whether or not there is a deep meridional flow in the
opposite sense.}
\label{Fig:methanecirc}
\end{figure}

These results present a picture in which dry upper tropospheric
gas descends at high latitudes after being dried out by condensation of methane
in upwelling gas at low latitudes, with upwelling and descending
regions connected by meridional cells as illustrated in Fig.\
\ref{Fig:methanecirc}. The general nature of this circulation is
similar to that described by KT2009 except that they don't
provide for the existence of a methane condensation cloud, which seems
to us a necessary mechanism to dry out the ascending gas.  The high
latitude descending gas appears to be very dry and to descend
only a small distance below the original methane condensation level.
However, this descent should inhibit the formation of methane clouds
at these latitudes, and if any do form they would have to form at much
lower pressures.  When we use five layers, our layer 3, which used to
be found at the methane condensation level, is now found much deeper
in the atmosphere and thus must be composed of some other substance,
perhaps part of a somewhat extended H$_2$S cloud layer.

\section{Latitude dependence of cloud structure}

\begin{figure}\centering
\includegraphics[width=3.4in]{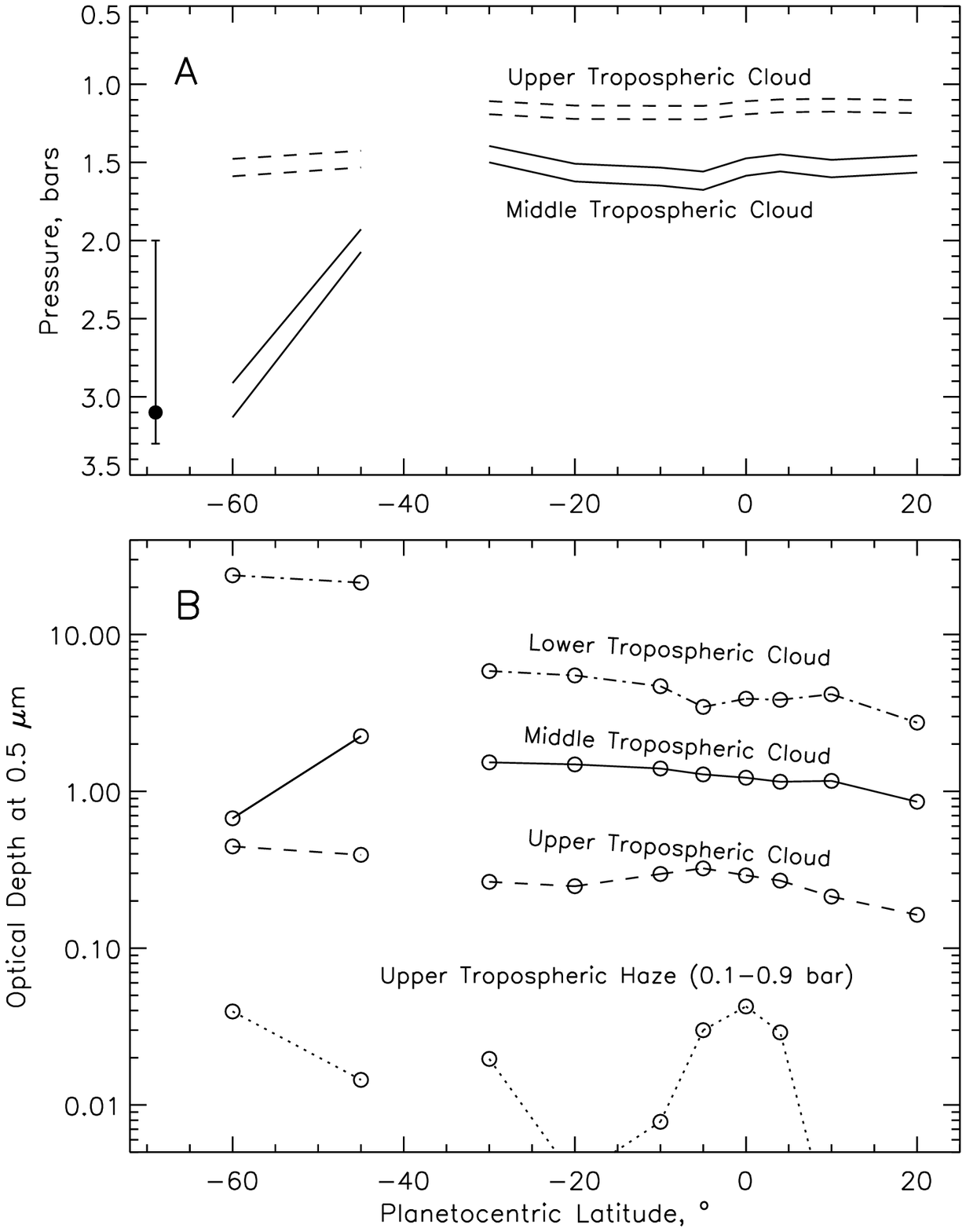}
\caption{Cloud model parameters as a function of latitude, from Table
\ \ref{Tbl:latfitsF1}.  The Upper Tropospheric cloud from 30\deg S to
20\deg N is consistent with a methane cloud. There appears to be no
methane cloud at 45\deg S and 60\deg S. The largest effect on the STIS
spectra is produced by the Middle Tropospheric Cloud. The filled
circle in A marks the top of the optically thick cloud of
\cite{Baines1995}, plotted at the sub-earth latitude of the
disk-integrated observations they analyzed.}
\label{Fig:latdep}
\end{figure}

Our cloud pressure and optical depth
parameters as a function of latitude are given in
in Table\ \ref{Tbl:latfitsF1} and key results are plotted in Fig.\
\ref{Fig:latdep}.  The results for latitudes from 30\deg S to 20\deg N
were all obtained using our F1 profile of methane and
temperature. For 60\deg S we used a methane profile depleted to 0.9\%
with a transition to 4\% between 3 and 4 bars.  At 45\deg S, we used a
depletion to 0.6\% with a transition to 4\% between 1.5 and 2 bars.
All three of these methane profiles are shown in Fig.\
\ref{Fig:depletions}A. Looking at the low-latitude region (30\deg S to
20\deg N), we see that the upper compact layer (the putative methane
layer, also known as the UTC or layer 3) changes pressure only slightly,
remaining consistent with methane condensation at the position of that
layer, while its optical depth increases slightly towards the equator
before beginning a significant decline into the northern hemisphere.
Somewhat more modulation is seen in the pressure of the dominant
middle tropospheric cloud (MTC, layer 4) and there is a substantial
declining trend in optical depth towards the north, though the rate
of decline decreases near the equator. Though these declines may
seem modest in the logarithmic plot, they are 7-10 times the typical
5\% fitting uncertainty.  The upper tropospheric haze
reaches a peak opacity near the equator, as noted by KT2009. 

The prominent bright band near 45\deg S that is observed in Uranus
images at wavelengths of intermediate absorption (0.1-0.4 /km-am, as
shown in Fig. 23 of KT2009) is also prominent in near-IR H- and J-band
images, as in Fig. 7 of \cite{Sro2007struc}. According to Fig.\
\ref{Fig:latdep}, this band is in a region where cloud pressures have
increased in both of the upper compact layers while the upper
tropospheric methane mixing ratio has decreased.  It is clear that the
upper compact layer at 45\deg S and 60\deg S cannot be the putative
methane ice layer, because the pressure levels are much too high to
permit condensation.  It is conceivable that in this region the upper
tropospheric cloud is actually what we called the middle tropospheric
cloud at low latitudes.  And the much deeper second compact layer may
be a new cloud layer formed by the upwelling produced by the lower
circulation branch in Fig.\ \ref{Fig:methanecirc}.  This would suggest
that the boundary between the upper and lower branches may be in the
1.5-1.7 bar region.  

The bright band contrast relative to 60\deg S is a result of the
deeper cloud being much deeper at 60\deg S and having a much lower
optical depth.  The contrast relative to lower latitudes results from
a combination of reduced methane opacity as well as the extra cloud
opacity in the 2-bar region, which apparently more than compensates
for the absence of the 1.2-bar cloud layer and the significantly
reduced opacity at 1.5 bars. At 60\deg S the middle tropospheric cloud
reaches about the same pressure as \cite{Baines1995} inferred for the
semi-infinite cloud in their model, which is heavily weighted towards high
latitudes.

\begin{table*}\centering
\caption{Fit parameters at 10 latitudes, using Model F1 profiles for all but 45\deg S and 60\deg S.}
\vspace{0.15in}
\begin{tabular}{r c c c c cc c c}
\hline\\[-0.1in]
 Centric & $P_{\mathsf{UTC}}$ & $P_{\mathsf{MTC}}$ &  $(d\tau/dP)$ & $(d\tau/dP)$ \\[0.05in]  
Latitude     &  bars            & bars & Strat.Haze &  UT. Haze
 & {\Large$\tau$}$_\mathsf{UTC}$   & {\Large$\tau$}$_\mathsf{MTC}$ 
   & {\Large$\tau$}$_\mathsf{LTC}$ &\chisq \\[0.05in] 
\hline\\[-0.1in]
20\deg N &1.185 &1.565 &0.230&0.0000 & 0.163& 0.86 & 2.74 &372 \\[0.05in] 
10\deg N& 1.175 &1.595 &0.334&0.0019 & 0.213& 1.17 & 4.16 &269\\[0.05in]
4\deg N & 1.180 &1.558& 0.148&0.0363 & 0.269& 1.15 & 3.83 &394\\[0.05in]
0\deg   & 1.192 &1.585& 0.185&0.0531 & 0.291& 1.22 & 3.89 &351 \\[0.05in]
5\deg S & 1.225 &1.676& 0.158&0.0375 & 0.322& 1.28 & 3.45 &342 \\[0.05in]
10\deg S& 1.224 &1.649& 0.280&0.0098 & 0.297& 1.40 & 4.68 &281\\[0.05in]
20\deg S& 1.222 &1.622& 0.262&0.0043 & 0.248& 1.48 & 5.48 &282\\[0.05in]
30\deg S& 1.192 &1.500& 0.075&0.0246 & 0.265& 1.53 & 5.85 &311\\[0.075in]
\hline\\[-0.1in]
45\deg S& 1.533 &2.074& 0.234&0.0181 & 0.393& 2.24 & 21.4 &435\\[0.05in]
60\deg S& 1.589 &3.131& 0.206&0.0493 & 0.445& 0.67 & 23.8 &444\\[0.05in]
\hline\\[-0.1in]
\end{tabular}\label{Tbl:latfitsF1}
\parbox{4.3in}{Note: Fixed parameters (except at 45\deg S and 60\deg
S) included the lower cloud base pressure (5 bars), the particle
radius of the upper tropospheric cloud (1.2 \mum). At 45\deg S and 60\deg S, we used
$r_{UTC}=1.1$ and 1.34 \mum, and $P_{LTC}=8$ and 11 bars, respectively.}
\end{table*}

\section{Discussion}

\subsection{Occultation failure to detect the middle tropospheric cloud}
Because the occultation probes to higher pressures than the pressures
we infer for the dominant middle tropospheric cloud, one might wonder
why the occultation did not detect this layer if it is indeed compact
and created by condensation? That turns out not to be a contradiction.
If the mixing ratio of the condensible is low enough, then the change
in molecular weight might actually be very small, and the resulting
refractivity change might be too small to be detected.  For example,
for H$_2$S to form a cloud base near 1.7 bars its volume mixing ratio would
need to be $\sim$10$^{-7}$ ($\sim$0.003 times the solar mixing
ratio). The maximum change in molecular weight induced by condensation
of that much H$_2$S would then be about
3$\times 10^{-6}$, which would be completely undetectable in the
occultation measurements, even if it occurred in a thin layer.  But
could such a cloud actually produce enough condensible material to
produce the needed reflectivity of the cloud?  The answer is yes. 
If half the H$_2$S gas in a 2-km thick layer were to condense
into 1-\mum droplets, we estimate that the
resulting cloud would have about 1 optical depth. Convective
processing of additional gas could increase its opacity to
much higher values.  NH$_3$ is a much less suitable candidate
because its corresponding mixing ratio would be three orders of
magnitude smaller as would the mass available for cloud
particle formation.

\subsection{Effects of para-hydrogen variations}\label{Sec:noneqh2}

\begin{figure*}\centering
\includegraphics[width=2.5in]{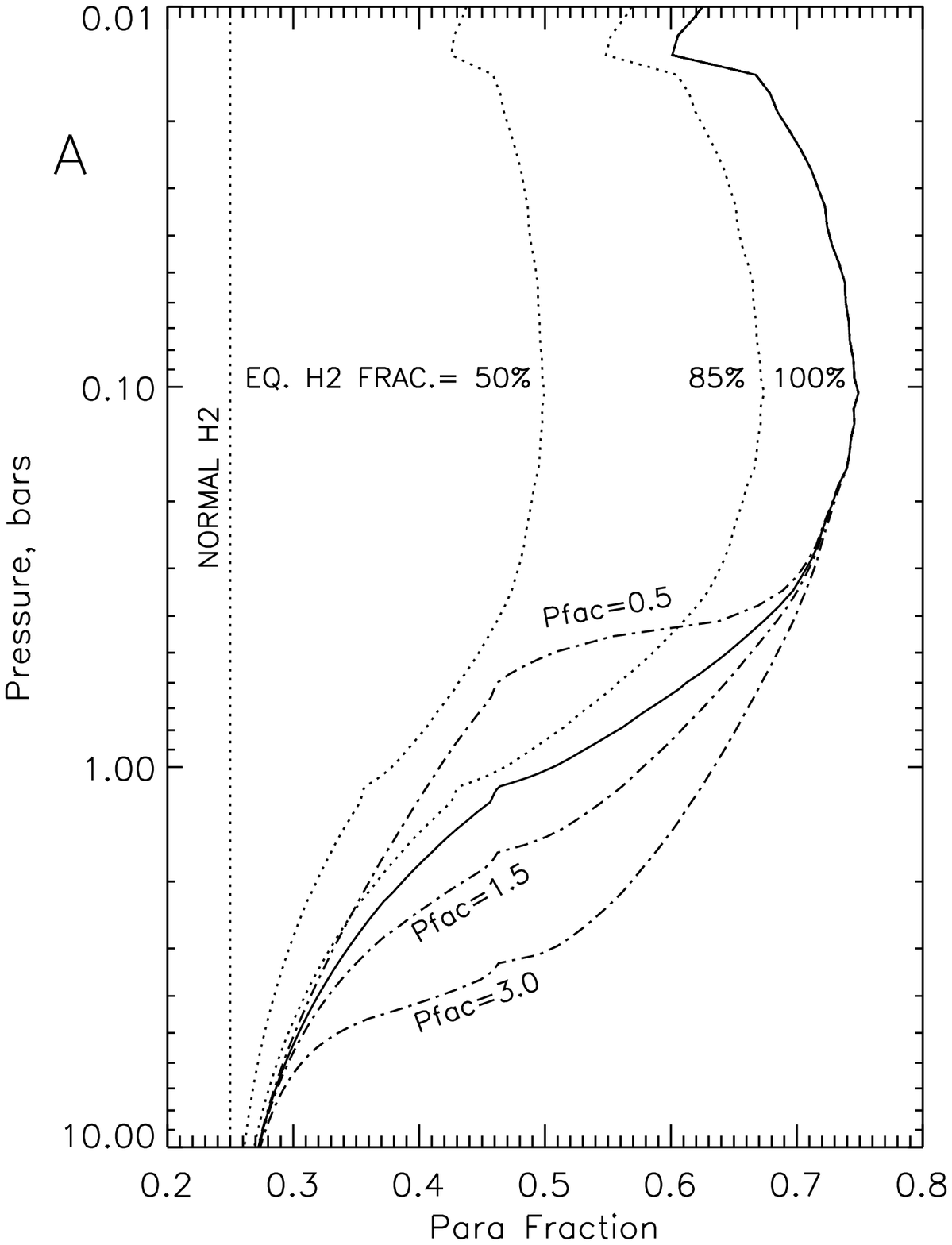}
\includegraphics[width=2.5in]{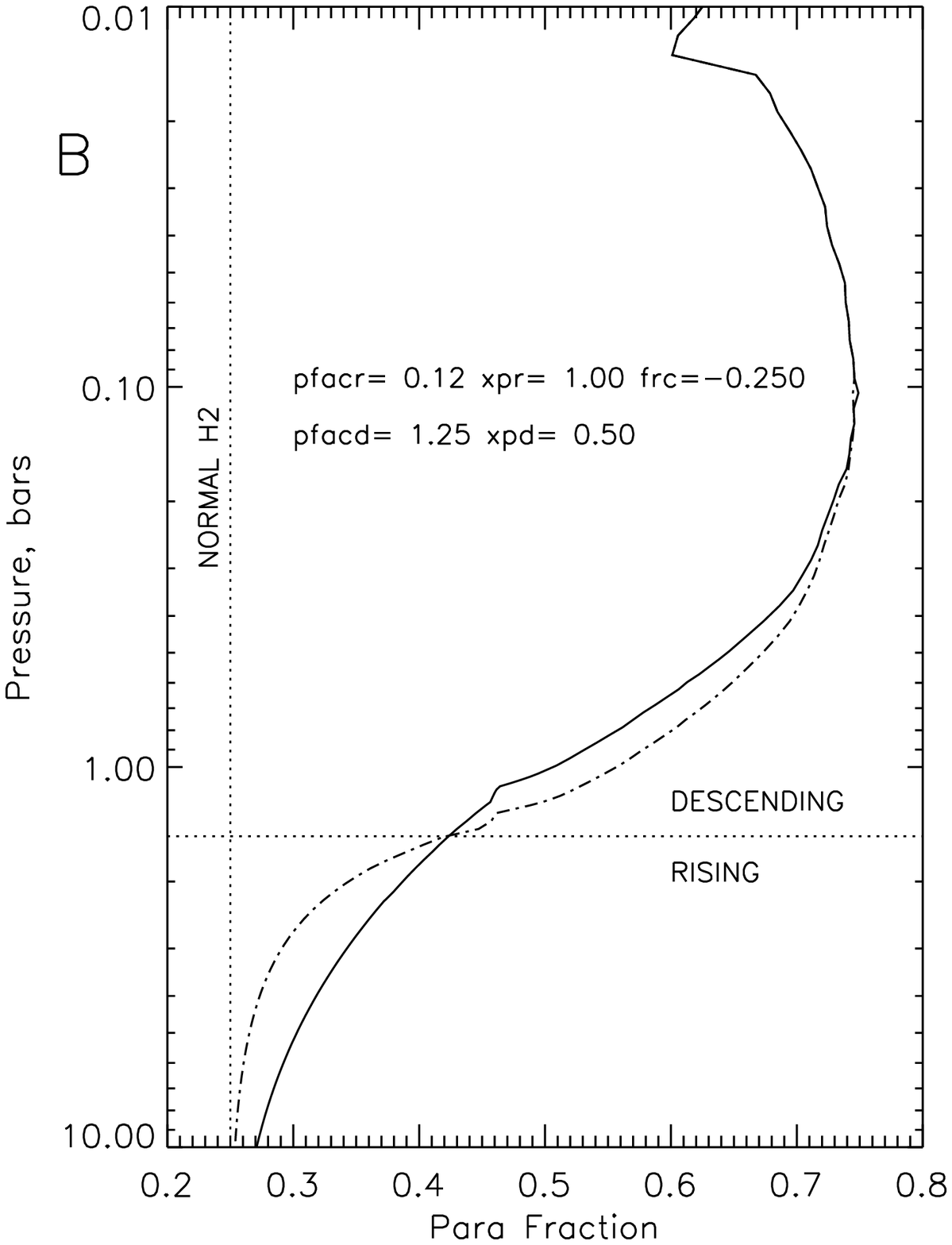}\par
\caption{A: Profiles of para hydrogen assuming local thermodynamic equilibrium (solid),
mixtures of equilibrium and normal hydrogen (dotted), and vertically
displaced hydrogen before new equilibrium conditions are established (dot-dash).
B: Example of a mixed rising-descending profile (dot-dash) compared to the
equilibrium profile.}
\label{Fig:parafractions}
\end{figure*}

All our calculations described so far have assumed equilibrium
hydrogen, for which the para and ortho fractions are in local
thermodynamic equilibrium. Since these two forms of hydrogen have
different spectral features, this assumption deserves some
consideration.  The equilibrium para faction for our F1 profile is
displayed in Fig.\ \ref{Fig:parafractions}, along with several
non-equilibrium curves, which we discuss in the following. In the deep
atmosphere high temperatures produce a para fraction of 0.25 (referred
to as normal hydrogen), but above the methane condensation level the
para fraction increases dramatically.  By vertically mixing deep
atmospheric gas up to the cloud level and higher, a dramatic change in
the para fraction can be produced, if done on a time scale short
compared to the relaxation time. \cite{Baines1995} considered mixtures
of equilibrium and normal hydrogen, as might be produced by vertical
mixing of the high temperature para distribution with the locally
equilibrated distribution.  Their results were consistent with
mixtures that were no less than 85\% equilibrium (and thus 15\%
normal).  KT2009 found that the difference between 100\% and 85\%
equilibrium had a smaller effect than their uncertainties.  They also
stated that 50\% equilibrium hydrogen would shift their results toward
lower aerosol opacities and lower mixing ratios, but not change their
results about latitudinal variations unless the the equilibrium
fraction itself varied with latitude, which they argued against on the
basis of smooth latitudinal variations in the observed hydrogen
absorption regions.

\begin{figure}\centering
\includegraphics[width=3.2in]{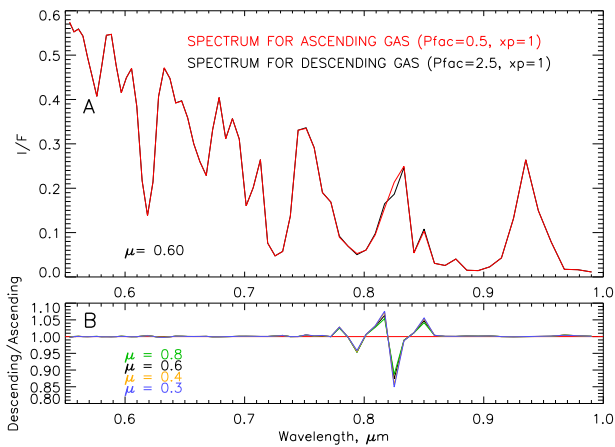}
\caption{A: Model spectra computed for ascending (red) and descending (black)
gas profiles at view angle cosine =0.6. B: Ratio of the two spectra at four
different view angle cosines.}
\label{Fig:paraeffect}
\end{figure}

Mixtures of equilibrium and normal hydrogen cannot produce the very
different para fraction profiles that can be produced by the
descending or rising branches of our putative methane circulation. For
convenience, we parameterize the rising or descending gas by Pfac,
which is the factor by which pressure at the midpoint of the
equilibrium profile is displaced. In Fig.\ \ref{Fig:parafractions}A,
we show para fractions that result when gas is vertically transported
upward (Pfac=0.5) or downward (Pfac=1.5 and Pfac=3.0), assuming that
the transport time is much less than the relaxation time to reach
local equilibrium. Note that mixtures of equilibrium and normal
hydrogen always decrease the para fraction, while vertical gas
transport can dramatically increase or decrease the local para
fraction.  
How changes in the 
para profile affect the Uranus spectrum can be seen in Fig.\
\ref{Fig:paraeffect} for two profiles of the type shown in Fig.\
\ref{Fig:parafractions}A: a descending gas case (Pfac=2.5) and an
ascending gas case (Pfac=0.5).  The effect is local, mainly between
0.77 and 0.86 \mumx, and is roughly the same percentage over a wide
range of view angles, reaching a maximum difference of about 15\%.
For the case shown in Fig.\ \ref{Fig:parafractions}B, where there is
descending gas above 1.5 bars, and ascending gas below (deeper than)
1.5 bars, which is more consistent with high-latitude motions shown in
Fig.\ \ref{Fig:methanecirc}, the spectral difference from the
equilibrium case is only a few percent, and not detectable within
current uncertainties.  The fact that we did not find evidence in the
spectra for significantly altered para profiles is consistent with
either such a combined upwelling/down-welling profile or no
significant departure from equilibrium hydrogen. However, this issue
may be worth further investigation, especially if methane absorption
coefficients become better known.  Part of the problem making use of
this constraint is that we are working with a part of the spectrum
where methane absorption is weak and thus currently is significantly
uncertain.

\subsection{Plausibility of a reduced He VMR on Uranus}

The Voyager Infrared Radiometer Interferometer and Spectrometer (IRIS)
and Radio Science Subsystem (RSS) together yielded a determination of
Jupiter's He/H$_2$ ratio equal to 0.114$\pm$0.025, which is a weighted mean
of results by \cite{Gautier1981} and \cite{Conrath1984}.  However, the
in situ measurements of the Galileo Probe contradicted this value,
instead obtaining 0.157$\pm$0.004 \citep{VonZahn1998} and
0.156$\pm$0.006 \citep{Niemann1998}.  \cite{Conrath2000} reviewed
possible explanations for this discrepancy and ruled out IRIS
calibration errors, but were unable to identify any plausible error in
the occultation measurements or analysis, although a detailed error
analysis of the occultation measurements was not conducted. This
discrepancy also cast suspicions on the very low He/H$_2$ ratio
obtained by IRIS-RSS analysis for Saturn, and led to an upward
revision of the Saturn ratio from 0.034$\pm$0.024 to 0.11-0.16, based
on a rather uncertain IRIS-only analysis \citep{Conrath2000}.  This
also raises questions about the occultation results for Uranus,
suggesting perhaps that the IRIS-RSS determined ratio is actually too
low for Uranus, rather than our suggestion that the ratio is too
high. But an even higher ratio, with the same refractivity profile
would lead to very serious amplification of inconsistencies with
near-IR and CCD spectra of Uranus that have already been noted.
Furthermore, because the specific error that caused prior
discrepancies is not known, it is not possible to predict with any
confidence how it might affect the Uranus analysis, or even what
direction it might take. The change in refractivity per molecule that
results from changing the He VMR from 0.15 to 0.11, as
computed from Eq.\ \ref{Eq:refrac} is from 4.496$\times 10^{-18}$ /cm$^3$ to
4.647$\times 10^{-18}$ /cm$^3$, 
which is a change of 3.3\%. We do not know if an
error of this magnitude is even remotely conceivable for the
occultation measurements.

\begin{figure}\centering
\includegraphics[width=3.2in]{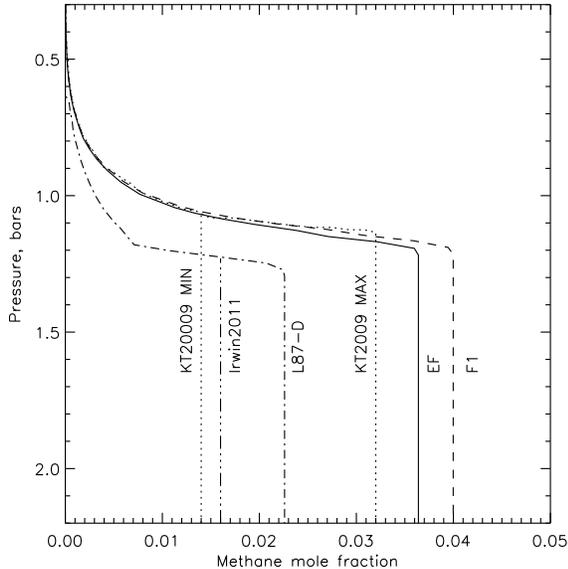}
\caption{Comparison of methane profiles from our EF and F1 models with those used by
KT2009, L87, and \cite{Irwin2011}, the latter following the L87 profile above the point
of intersection of their assumed deep 1.6\% VMR with that profile.}
\label{Fig:ch4comp}
\end{figure}

\subsection{The reasons for higher cloud pressures obtained from Near-IR observations}

The optical depth of the methane cloud is relatively small and the
influence of the middle tropospheric cloud is much more significant
because of its greater opacity. That and the limited vertical
resolution inherent in all the spectral observations, will lead
inversion methods such as the NEMESIS routine used in by Irwin and
colleagues \citep{Irwin2010Icar, Irwin2011} to find peak opacity
contributions at pressures deeper than that of the methane cloud.  In
addition, these and most prior modelers of near-IR spectra assumed
profiles of methane with much smaller column amounts of methane, often
by a factor of two or more (Fig.\ \ref{Fig:ch4comp}), which leads to
much higher pressures for the level of significant aerosol scattering.
The dependence of derived pressure on assumed methane was clearly
demonstrated by \cite{Sro2007struc} and by \cite{Irwin2010Icar}. As
noted previously, when Irwin et al. switched from Model D to the Model
F profile, the pressure of their peak opacity concentration moved from
2.5 to 1.7 bars.  Since our Model F1 has more column abundance of
methane down to the 1.2 bar level than even the L87 Model F, we expect
similar pressures would be obtained using our recommended profile.


\section{Summary and Conclusions}

After reanalysis of the L87 radio occultation profiles of refractivity
vs altitude, and fitting compact cloud layer models to STIS spectra as
corrected and calibrated by KT2009, we reached the following main
conclusions:

\begin{enumerate}

\item By decreasing the stratospheric He mixing ratio from its
  nominal value of 0.15 by 1-1.3 times its uncertainty it is possible
  to achieve methane saturation within the layers suspected to have
  condensation and to achieve increased methane humidities above the
  condensation level, even exceeding the high values adopted by KT2009
  at low latitudes, which are inconsistent with the original L87 profiles. The
  maximum deep mixing ratio that we could obtain within reasonable
  physical constraints was 4.88\% (for our model G).

\item A five-layer cloud model in which the bottom two diffuse layers of the KT2009
model are split into three compact layers, when constrained by STIS spectra
at 5\deg S, yield best-fit pressures for the top compact layer in excellent agreement
with the location of the occultation cloud layer for profile models with deep
\chf mixing ratios between 3.2 and 4.5\%, with the best compromise fit being obtained
at 4\% (for our Model F1), although this fit is not significantly better
than for the other models in this range.

\item As judged by fitting errors at 0.825 \mumx, where H$_2$ CIA exceeds
methane absorption, the best compatibility between methane and
H$_2$ CIA is obtained for a methane mixing ratio of 4.5\% with an
uncertainty range of about 0.7\%.

\item When we consider constraints of upper cloud pressure, fit error
at 0.825 \mum, overall fit quality, and the uncertainty limits of the
\cite{Conrath1987JGR} helium abundance, our best compromise estimate for
the deep methane mixing ratio at 5\deg S is 4.0$\pm$0.5\%, and the
corresponding preferred vertical temperature and methane profiles
are embodied in our F1 Model.  

\item Our five-layer cloud model, using the EF and/or F1 profiles, can
also provide excellent fits between latitudes of 30\deg S and 20\deg
N, with relatively latitude-independent cloud pressure boundaries, but
generally decreasing optical depths from south to north. The main
exception is the upper tropospheric haze, which shows a strong peak
near the equator, as noted by KT2009. The lower tropospheric cloud is
found to have the greatest opacity, but the 1-2 optical depth (at 0.5
\mumx) middle tropospheric cloud (in 1.5-1.7 bar range) has the
most significant effect on the observed CCD spectrum.  The pressure of
the 0.15-0.3 optical depth methane cloud layer did not vary much with
latitude, staying within 2\% of its 1.2 bar mean.

\item Trying to fit our model at high southern latitudes made it clear
that the methane mixing ratio profile could not remain the same as we
used in the low latitude observations.  Not only did overall fit quality
deteriorate, but the fit quality at 0.825 \mum wavelength of H$_2$ CIA
got suddenly very bad, and the direction and size of the error at that wavelength
indicated a significant lowering of the methane mixing ratio, as first
pointed out by KT2009.

\item We created depleted methane profiles in which depletions reached
a limited depth at which point the mixing ratio transitioned to the
same deep value of 4\% used at low latitudes. We did not expect a
depletion at great depths because the horizontal density gradients
that would be generated would induce problematic vertical wind shears.
Overall and 0.825 \mum fit quality was found to be minimized with
mixing ratios less than 1\% and depletion depths of 3-4 bars at 60\deg S
and only 1.5-2 bars at 45 \deg S.

\item Creating the observed methane depletion is plausibly the result
of upwelling at low latitudes, where the Uranus atmosphere is dried out by
condensation of cloud particles, and subsequent poleward transport to
and descent of the dried-out atmosphere at high latitudes down to levels of a
few bars, with greater descent nearer to the pole. This would inhibit
methane condensation at high latitudes and perhaps help to explain the
failure to detect any discrete cloud features from 45\deg S to the
south pole. 

\item The failure of modelers of near-IR spectra to detect a methane cloud at the 
proper location (where the occultation found a sudden change is refractivity) is
due to two effects: one is that most modelers used mixing ratio profiles providing
too little methane absorption and the other is that the methane cloud layer has a significantly
lower opacity than the main cloud layer that peaks in the 1.4-1.7 bar region, making
it hard to resolve its separate contribution.

\item The failure of the occultation measurements to detect the deeper
and more significant middle tropospheric cloud layer that so prominently affects both
near-IR and CCD spectra is not at all surprising if it is created by
condensibles at very low mixing ratios. In that case the condensation would not produce
a detectable change in refractivity.

\item Although it is not possible for the STIS spectra alone to clearly distinguish
between vertically extended models, such as that of KT2009, and compact layer
models of the type we presented here, there is a strong distinction in physical
processes involved.  Our model is consistent with cloud formation by condensation, while
the KT2009 model presents a picture in which a vertically extended haze of stratospheric
origin provides all the particulate opacity, much like the vertically extended
haze on Titan. We prefer the condensation model because it provides consistency with
the occultation observations and provides a mechanism for depleting methane
from the upper troposphere at high latitudes.  

\end{enumerate}

\noindent 
Many of these results are dependent on models of methane absorption in spectral
regions where absorption is relatively weak and somewhat uncertain. Some 
of the absorption coefficients were adjusted by KT2009 to obtain better
overall fits to the observations, which might tend to favor the particular
models they were using in their analysis.  As we obtain more accurate
information about methane absorption that is independent of the spectral
observations we are trying to interpret, these conclusions may need to
be modified.

The view of Uranus' north polar regions at the beginning of 2011 is
now adequate to address an important issue with new observations.  The
issue is whether the depletion of methane observed at southern high
latitudes is also present at northern high latitudes, and if not
currently depleted, whether it will become depleted as Uranus seasons
proceed. There is some evidence suggesting high northern latitudes
might not be depleted: discrete cloud features have been observed in
2007 Keck images between 45\deg N and 74\deg N \citep{Sro2009eqdyn},
while no discrete cloud features have been seen at corresponding
southern latitudes. The lack of southern discrete features is
correlated with the inferred descent of dry gas at those latitudes,
suggesting that the descending gas and resulting methane depletion may
not be present at high northern latitudes.  The question of whether
methane is currently depleted at high northern latitudes on Uranus can
be answered unequivocally by new STIS spectral measurements.

\section*{Acknowledgments}

This research was supported by NASA Outer Planets Research Grant
NNG05GG93G and Planetary Atmospheres Grant NNX09AB67G.  We thank
E. Karkoschka and M. Tomasko for making their calibrated STIS data
cubes available to the community. We thank two anonymous reviewers for
constructive comments.



\end{document}